\documentclass[draft]{agujournal2019}
\usepackage{url} 
\usepackage{lineno}
\usepackage{soul}
\usepackage{amsmath,amssymb,stix}
\draftfalse

\journalname{Journal of Advances in Modeling Earth Systems (JAMES)}

\begin{document}

%
%


\title{Rare event sampling for moving targets: extremes of temperature and daily precipitation in a general circulation model}

%
%




\authors{Justin Finkel\affil{1,2} and Paul A. O'Gorman\affil{1}}


\affiliation{1}{Department of Earth, Atmospheric and Planetary Sciences, Massachusetts Institute of Technology}
\affiliation{2}{Department of Geophysical Sciences and the Data Science Institute, University of Chicago}




\correspondingauthor{Justin Finkel}{jfinkel@uchicago.edu}



\begin{keypoints}
\item Extreme weather risk is highly uncertain, but can be estimated more accurately by targeted rare event sampling. 
\item Rare event algorithms are challenged by short time scales of weather events which limit ensemble diversity.
\item Optimally timed perturbations enable sped-up probability estimates of precipitation and heat extremes in an aquaplanet climate model. 
\end{keypoints}

%
%

%
%


\begin{abstract}
Extreme weather events epitomize high cost: to society through their physical impacts, and to computer servers that simulate them to assess risk and advance physical understanding. It costs hundreds of simulation years to sample a few once-per-century events with straightforward model integration, but that cost can be much reduced with \emph{rare event sampling}, which nudges ensembles of simulations to convert moderate events to severe ones, e.g., by steering a cyclone directly through a region of interest. With proper statistical accounting, rare event algorithms can provide quantitative climate risk assessment at reduced cost. But this can only work if ensemble members diverge fast enough. Sudden, transient events characteristic of Earth's midlatitude storm track regions, such as heavy precipitation and heat extremes, pose a particular challenge because they come and go faster than an ensemble can explore the possibilities. Here we extend standard rare event algorithms to handle this challenging case in an idealized atmospheric general circulation model, achieving $\sim5-10$ times sped-up estimation of long return periods for extremes of surface temperature and daily precipitation (e.g., a return period of 150 years from 20 years of simulation). The algorithm, called TEAMS (``trying-early adaptive multilevel splitting''), was developed previously with a toy chaotic system, and relies on a key parameter---the advance split time---which may be estimated based on simple diagnostics of ensemble dispersion rates. The results are promising for accelerated risk assessment across a wide range of physical hazards using more realistic and complex models with acute computational constraints. 
\end{abstract}

\section*{Plain Language Summary}
Climate hazards are largely felt not through global mean temperature, but through extreme weather events, which are dangerous not only for their physical severity but also for their rarity: by definition, they are very difficult to anticipate and prepare for. The same characteristic makes risk assessment a very hard statistical problem. Numerical simulations can be used to augment small sample sizes, but at great computational cost. Rare event algorithms offer a novel way to ``steer'' simulations towards the extremes to do targeted risk assessment at reduced cost, but this can be challenging when the events under study are transient in nature, such as passing rainstorms and heat extremes in Earth's midlatitude region. This paper presents a successful application of a rare event algorithm to such transient extremes in an idealized model of Earth's atmospheric circulation, building on previously published results that used a simpler toy model of spatial chaos. The core of the method is to select the right time to perturb the simulations, and the fact that the method generalizes is a promising sign that it can scale to even more complex, realistic models.

%
%

%


%
%
%
%

\section{Introduction}
The highest-impact extreme weather events are those that occur so seldom as to catch communities---cities, ecologies, and scientists alike---surprised and unprepared \cite{Sillmann2017understanding}. Even with physically accurate numerical models capable of simulating extremes, running them long enough to collect ample statistics can be prohibitive. A key innovation to close this gap is \emph{rare event sampling}, a protocol which steers ensembles of simulations towards the extremes by repeated perturbation, pruning, and cloning steps, while adjusting probability weights on ensemble members to compensate for preferential sampling and thus enable unbiased statistical estimation. Originally developed for nuclear physics simulation \cite{Kahn1951estimation}, rare event algorithms have been specialized and developed for molecular dynamics \cite{Zuckerman2017weighted}, reliability engineering \cite{Huang2016assessing,Sapsis2020output,Uribe2021cross,Zhang2022koopman}, and climate science \cite<e.g.,>[]{Ragone2018computation,Wouters2016rare,Webber2019practical}. Rare event algorithms are attractive for being agnostic to the model: importantly, they can operate on models grounded in physics and potentially could also be applied to faster, data driven models with the alluring possibility of generating abundant extreme events at will \cite{Mahesh2024hensone,Mahesh2024henstwo}.

Yet there remain some methodological roadblocks to the broad deployment of rare event algorithms across different models and different rare events. This paper addresses one such roadblock: an overlap of timescales between ensemble dispersion and event duration. If the event duration is too short, the ensemble has too little time to spread out and sample the tails before the event is over. This is not a problem for long-lasting, spatially extended events such as hot or rainy \emph{seasons}---defined by large \emph{seasonal mean} temperature or precipitation amplitudes. Such events are already a successful application for rare event algorithms \cite{Ragone2018computation,Wouters2016rare}, as multiple successive rounds of ensemble splitting can fit into a single season, with enough time for dispersion between each round, so that extreme anomalies can be achieved by essentially chaining together a sequence of moderate anomalies. But transient events of much shorter duration don't yield so easily; na{\"i}vely applying the same perturbation protocol simply results in disappointing replication of the same moderate extreme again and again, without meaningful exploration into the far tails \cite{Lestang2020numerical,Rolland2022collapse,Finkel2024bringing}. This is a major limitation given that transient cyclones and anticyclones can bring heavy rain and temperature extremes that are among the most impactful extreme events for society.

We developed a simple remedy to this problem by adapting the classical Adaptive Multilevel Splitting (AMS) algorithm \cite{Cerou2007adaptive,Lestang2018computing} in which a level of extremity is progressively raised and only ensemble members reaching that level are retained and split at a level crossing. We modified AMS to perturb simulations in advance of the level crossing, thus giving more time for ensemble members to separate before reaching an extreme event. Splitting in advance requires an additional  acceptance/rejection step that we include through the formalism of subset simulation \cite{AU2001estimation}. The resulting algorithm, TEAMS (``trying-early adaptive multilevel splitting''), introduces a key hyperparameter, the \emph{advance split time} (AST), which determines when to split the simulation relative to the event for an optimal balance of exploration (with high risk of rejection) and exploitation (with low risk of rejection but limited rewards).  TEAMS draws inspiration from the related approach of \emph{ensemble boosting} \cite{Gessner2021very,Gessner2022physical}  which has recently been extended to include probability estimates \cite{BloinWibe2025estimating,Finkel2025boosting}. Ensemble boosting differs from TEAMS in that it perturbs before the extreme events in an existing long simulation, and perturbs them all the same number of times without sub-selection based on a level-raising protocol.  In \citeA{Finkel2024bringing} we demonstrated TEAMS on the Lorenz-96 system, a relatively simple model of spatiotemporal chaos that nevertheless captures the essence of baroclinic waves. Our main contribution here is to demonstrate a successful use of TEAMS on an actual climate model, albeit an idealized one, to sample short-timescale events, namely high surface temperatures and daily precipitation rates.     

This paper is organized as follows. Section \ref{sec:model} briefly specifies the physical model, a general circulation model (GCM) in an aquaplent configuration, emphasizing two modifications of reduced resolution for computational efficiency and the addition of stochastic parameterization. Section \ref{sec:algorithm} outlines the rare event algorithm TEAMS, emphasizing the most recent modifications of how rejection is handled and the halting criteria. Section \ref{sec:results} shows the results of applying TEAMS: efficiency gains in calculating long return periods (100 years and longer), and the generation of corresponding dynamical samples. Section \ref{sec:conclusion} concludes with a summary and outlook on further avenues of development.

\section{The physical model}\label{sec:model}

We use an idealized GCM based on the GFDL spectral model and similar to that developed in \citeA{Frierson2006gray} with slight modifications as in \citeA{OGorman2008hydrological}. A spectral dynamical core integrates the primitive equations, with a lower boundary condition consisting of a slab ocean (aquaplanet) that is shallow, well-mixed, and energy-conserving (not fixed-temperature). Insolation is fixed to an average distribution, with no seasonal or diurnal cycle. A two-stream gray radiation scheme is used, with a prescribed distribution of longwave optical depth. We turn off the convection parameterization, so that condensation of water vapor occurs only at the large scale (grid box size), as was found to be adequate for midlatitudes by \citeA{Frierson2006gray}. Turbulent diffusivities are smoothed in time following \citeA{Anderson2004new}.

We make two further modifications for this rare event sampling demonstration. To enable computational efficiency, we reduced the horizontal spectral resolution to T21, meaning a triangular truncation of spherical harmonics with maximum wavenumbers $21$ in both zonal and meridional directions \cite{Krishnamurti2006introduction}. 
We also reduce the temporal and vertical resolution, using a 40-minute timestep and six $\sigma$-levels in the vertical (half levels at $\sigma=$ 0.0, 0.0343, 0.15, 0.4, 0.7, 0.966, 1.0), where $\sigma$ is pressure normalized by surface pressure.  We also present some limited results at a higher horizontal resolution of T42, with 30 vertical levels and a 10-minute timestep. The simulations were performed with MPI on four Intel Xeon cores, completing 60 days of simulation in roughly 20 seconds at T21 resolution, or 15 minutes at T42 resolution. All results are at the default resolution of T21 unless otherwise noted.

The second modification is to introduce the randomness needed to induce variability between ensemble members. Other rare event sampling methods \cite<e.g.,>[]{Ragone2018computation,BloinWibe2025estimating,Abbot2021rare} used single-time perturbations. Here, in line with the continuous-time forcing used in \citeA{Finkel2024bringing}, we implemented a stochastic parameterization scheme known as stochastically perturbed parameterized tendencies (SPPT). SPPT was developed in numerical weather prediction to enhance ensemble spread to more likely capture the observed evolution \cite{Palmer2009stochastic,Berner2009spectral,Berner2015increasing}, and here we can use it to discover unlikely paths towards extremes. Our implementation of SPPT closely follows the specification in \citeA{Palmer2009stochastic}, which contains further details and background. In brief, SPPT randomly perturbs the total parameterized tendencies (that is, contributions from large-scale condensation, vertical turbulent diffusion, and radiation) of horizontal winds, humidity, and temperature. The perturbation acts every timestep through a multiplicative factor of $1+r_{\mathrm{SPPT}}(x,y,z,t)$, where $r_{\mathrm{SPPT}}(x,y,z,t)$ is a random spatiotemporal pattern whose spherical harmonic modes each evolve as an independent red noise process. There are three tunable parameters: characteristic autocorrelation timescale $\tau_{\text{SPPT}}$, length scale $L_{\text{SPPT}}$ which specifies how quickly amplitude drops off with wavenumber, and an overall multiplier $\sigma_{\text{SPPT}}$. To prevent unrealistically large fluctuations, $r_{\text{SPPT}}$ is clipped to the range of $\pm2$ standard deviations. Sensitivity analysis led us to select $L_{\mathrm{SPPT}}=500$ km, $\tau_{\mathrm{SPPT}}=6$ hours, and $\sigma_{\mathrm{SPPT}}=0.3$. We gave especially careful consideration to $\sigma_{\mathrm{SPPT}}$, the overall noise amplitude, because the analogous stochastic forcing strength that \citeA{Finkel2024bringing} used on Lorenz-96 was found to strongly affect optimal advance split time and the extent to which TEAMS improved on AMS. The choice of $\sigma_{\mathrm{SPPT}}=0.3$ will be justified in Fig. 1, and is quite similar to the moderate-amplitude experiments in \citeA{Palmer2009stochastic}. 

We use this computationally efficient GCM because it accommodates the large ensemble sizes and parameter tuning experiments needed for development and testing of rare-event sampling strategies. Our aim is to demonstrate a novel methodology more than a particular scientific conclusion, and for this purpose a lower rung on the model hierarchy \cite{Held2005gap} take on greater value. The same idealizations (such as zonally symmetric boundary conditions) that make this model attractive for extensive parameter sweeps, as in \citeA{OGorman2008hydrological} and \citeA{OGorman2009scaling}, also make it well-suited for rare event algorithm development. At the same time, even the coarse model is physically realistic enough that the insights learned here should transfer to more realistic models. 

\begin{figure}
    \centering
    \includegraphics[trim={0cm 0cm 3cm 0cm},clip,width=0.98\linewidth]{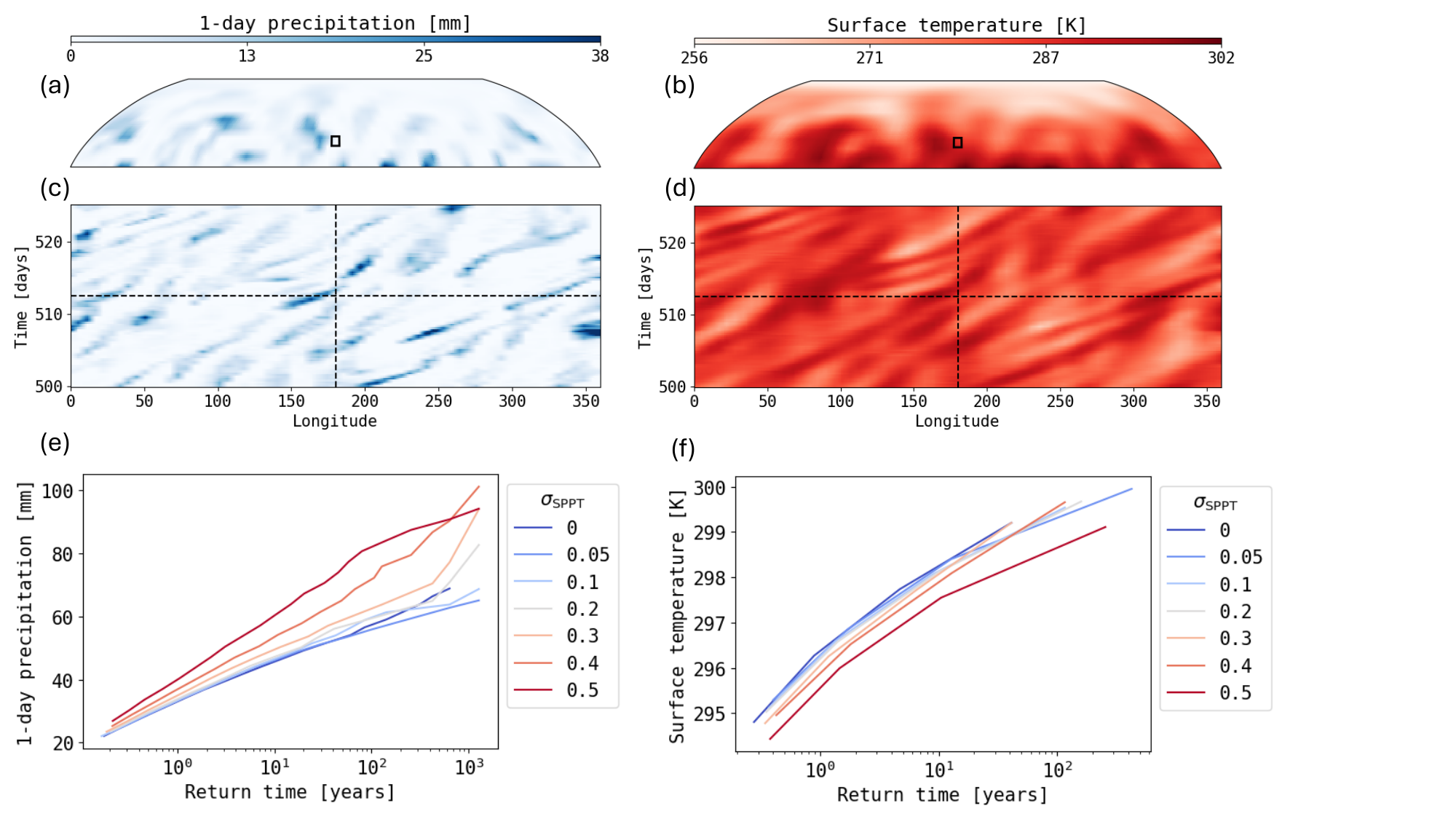}
    \caption{Simulated precipitation and surface temperature fields and their return levels at T21 resolution. After a spin-up of 500 days, the aquaplanet GCM produces physically plausible large-scale storm track dynamics: a sequence of extratropical cyclones and anticyclones bringing packets of precipitation (a) and temperature fluctuations (b), propagating eastward with lifetimes of $\sim5$ days (Hovm\"oller diagrams in c and d). We select a target region (one grid cell marked by a black square in (a,b)) to fall at $45^\circ$N, near the latitude of maximum mean precipitation, and a longitude of $180^\circ$E (which is arbitrary because climatological statistics are zonally uniform). Horizontal and vertical dashed lines in (c,d) indicate the timing of the snapshot and the target longitude. Panels e,f show return level vs. return period plots of both targets, local precipitation and temperature, for a range of values of the SPPT forcing strength $\sigma_{\mathrm{SPPT}}$. The return levels vary only moderately for $\sigma_{\mathrm{SPPT}}\lesssim 0.3$ and start deviating substantially for larger values, which is why we adhere to $\sigma_{\mathrm{SPPT}}=0.3$ in panels (a-d) and hereafter.}
    \label{fig:maps_precip_temp}
\end{figure}

Fig. \ref{fig:maps_precip_temp} displays some characteristics of the surface temperature and precipitation fields produced by the GCM once it reaches statistical equilibrium after a spinup period. Throughout the paper, surface temperature refers to the surface air temperature evaluated at the lowest model level. Outputs from the GCM are six-hourly; temperature is instantaneous (noting there is no diurnal cycle) and precipitation is averaged over the previous day.
Despite the idealized setup and coarse resolution, the baroclinic waves of Earth's midlatitude storm track and associated precipitation and temperature variability are clearly visible in the model fields (Fig. \ref{fig:maps_precip_temp}a,b), which grow and decay over synoptic $\sim5$-day timescales (indicated by the Hovm\"oller diagrams in Fig. \ref{fig:maps_precip_temp}c,d). 
Our aim is to characterize---using rare event sampling---the extreme, local fluctuations in these fields at the storm track's center. We therefore fix a target latitude of $45^\circ$N and a target longitude of $180^\circ$E, taking the field value in a single grid cell $(\sim6^\circ)$ as the target variable. 
The choice of longitude is arbitrary due to the model's zonal homogeneity, but fixing a longitude simplifies the event definition and would be necessary anyway in Earth system models with zonal asymmetries.
Still, we take advantage of zonal homogeneity in computing ``ground truth'' statistics from long simulation by pooling together eleven longitudinal rotations in $30^\circ$ increments for more stable estimation with twelve times the data. 
Fig. \ref{fig:maps_precip_temp}(e,f) displays the long-term climate statistics of precipitation and temperature at the target location, revealing $\sigma_{\mathrm{SPPT}}\approx0.3$ to be near the upper limit of noise level that still avoids disrupting the deterministic model's statistics too severely. The effect of noise is to increase precipitation extremes but decrease mean temperatures.   
\citeA{Tagle2016temperature} found stochastic parameterization increased mean temperatures  
in the Community Atmosphere Model, and the different result for temperature found here may relate to the idealized GCM we use which does not include land or cloud radiative effects.

The results in Fig. \ref{fig:maps_precip_temp} at T21 resolution are based on a long run of 36,500 days (100 years, or 1200 years including longitudinal rotation) after spinup, which we refer to as a direct numerical simulation (DNS). For validation of TEAMS results shown later, we extended the dataset at $\sigma_\mathrm{SPPT}=0.3$ even further: at T21 resolution we generated 272.8 years of DNS with 30 longitudinal rotations for 8184 years total; and at T42 we generated 8.4 years of DNS with 30 longitudinal rotations for 253 years total. The data used for initializing TEAMS, on the other hand, is branched from the long DNS after spinup and integrated independently, with a different seed for each run of TEAMS, in order to avoid data leakage (see ``ancestor initialization'' in the algorithm described in section \ref{sec:algorithm}). 

\section{The TEAMS algorithm}\label{sec:algorithm}

\begin{figure}
    \centering
    \includegraphics[width=0.99\linewidth,trim={1cm 8cm 16cm 1cm},clip]{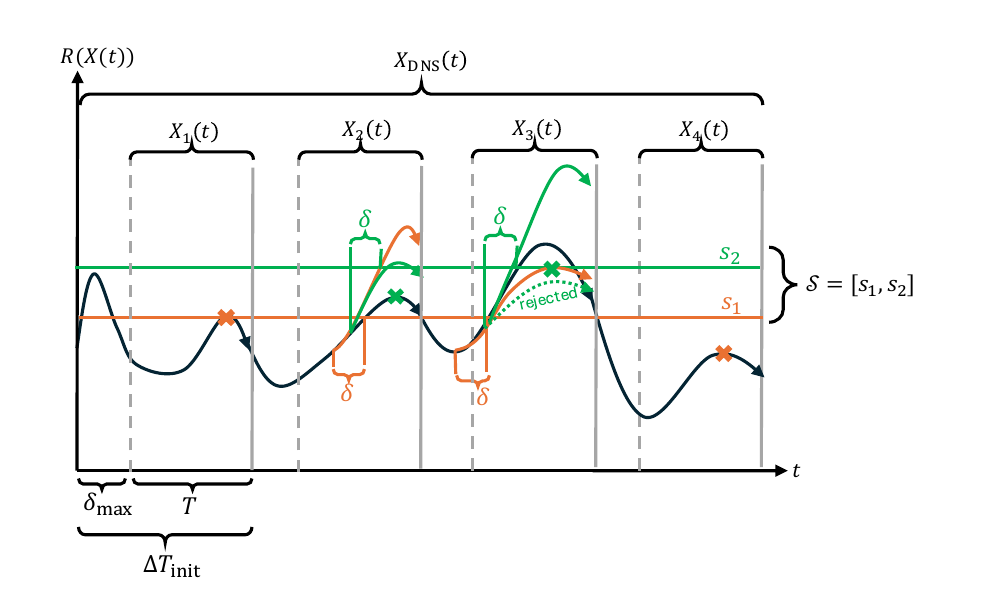}
    \caption{Schematic of TEAMS in a simplified setup with $N=4$ ancestors and only two level-raisings. 
    Orange elements pertain to the first culling and cloning step; green elements pertain to the second. Horizontal lines are levels $s$ set by the algorithm, and crosses denote maxima that are eliminated as a result of falling below the level. The black line is the DNS used to generate the ancestors. Splits occur a time $\delta$ before a level crossing. Severities are measured over a time horizon $T$ that follows a runway $\delta_\mathrm{max}$. See text for further details.}
    \label{fig:schematic}
\end{figure}

Let us briefly describe the TEAMS algorithm, following \citeA{Finkel2024bringing}. Along the way we delineate between generic parameter choices and those made in this study to target local temperature and precipitation extremes in the GCM. 
Readers interested primarily in the sampling results can skip to Section \ref{sec:results}. Fig. \ref{fig:schematic} serves as a visual reference for the key elements of the procedure. 

\begin{enumerate}
    \item\underline{Ancestor initialization}: Sample $N$ initial conditions $\{X_1(0),X_2(0),...,X_N(0)\}$ from the distribution of interest, denoted $\rho_0$. For us, $\rho_0$ is the distribution at statistical steady state, i.e., the limiting distribution of a very long GCM simulation. Other applications might restrict the initial conditions to specific phases of an oscillation (e.g., neutral El Ni\~no conditions) or, if a seasonal cycle is present, specific dates (e.g., June 1 conditions). For our study, we can extract the $X_n(0)'s$ as snapshots from a direct numerical simulation (DNS), which is branched from the DNS used for validation by changing the random seed for SPPT after spinup. Consecutive ancestral initial conditions are separated by a gap of $\Delta T_\text{init}=60$ days, chosen as roughly twice the \emph{error saturation} timescale (over which two branched simulations decorrelate).  
    
    \item\underline{Ancestor simulation}: Run the dynamics forward for a \emph{time horizon} $T$  from each ancestral initial condition. We set $T=35$ days, long enough to contain 1-2 events and ample time for a branched ensemble to decorrelate. The resulting dataset is $\{X_n(t):1\leq n\leq N, 0\leq t\leq T\}$, where we have re-indexed time for convenience. In our setup, no new simulation is necessary to create the ancestor simulations: we just use the DNS segments directly. We reserve  
    $\delta_\text{max}$ days before each initial condition and after the previous terminal condition as a `runway', where $\delta_\text{max}=25$ was chosen as an upper bound on the advance split times considered. The runway is needed for cases in which the split time occurs before the ancestral initial condition. Both the runway and the remaining time horizon of $T=\Delta T_\text{init}-\delta_\text{max} = 35$ days helps ancestors be more independent, which is not a strict requirement (they need only be identically distributed), but improves statistical robustness.  Thus, each ancestor simulation 
    takes the form
    \begin{align}
        X_n(t)=X_\text{DNS}\Big(\delta_\text{max}+(n-1)\Delta T_\text{init}+t\Big).
    \end{align}
    Fig. \ref{fig:schematic} depicts how the ancestor simulations are laid out relative to the DNS simulation.
    
    Assign each ancestor a probability weight $W_n=1$. Furthermore, initialize a set of \emph{active members} 
    \begin{align}
        \mathcal{A}=\{1,\hdots,N\}
        =:\{a_1,\hdots,a_A\}
    \end{align}
    with a size $A=N$, which will be modified by repeated culling and replenishment in following steps. Also initialize an empty list of \emph{severity levels} $\mathcal{S}=[]$, which will grow in the following steps.

    \item\underline{Culling}: Rank the active ensemble members $a\in\mathcal{A}$ by their \emph{severity}, $S_a=S(X_a)$ defined as the peak value over time of the \emph{intensity} $R_a(t)=R(X_a(t))$ which defines the target variable of interest. In our case, our outputs are six-hourly and  $R(X_a(t))$ is the precipitation (averaged over the preceding four snapshots $=$ one day) or surface temperature (measured at a single six-hourly snapshot) at the target grid box indicated in Fig. \ref{fig:maps_precip_temp}. Choose a number $K<A$ and cull the the $K$ least-extreme active members. We choose $K=\frac12N$, but one could also set $K$ as a constant number (commonly $K=1$, as in \citeA{Finkel2024bringing}) or some other fixed fraction of $N$ (in engineering applications, the related ``subset simulation'' algorithm commonly culls aggressively with $K\sim0.9N$ \cite{AU2001estimation}). We denote $s$ as the level that will be progressively raised as the TEAMS algorithm proceeds. We set $s$ equal to he $K$-th smallest severity such that it has an  estimated exceedance probability of $(N-K)/N$ (for us, $1/2$). Append the list of severity levels, $\mathcal{S}\leftarrow\mathcal{S}\cup[s]$. Remove the culled members from the active set, reducing its size to $A-K$, re-index its members accordingly to $\mathcal{A}=\{a_1,\hdots,a_{A-K}\}$, and reset the size $A$ to $A-K$.  

    \item\underline{Cloning}: Shuffle the active members in a random order, called the ``parent queue''. For the first parent $a$ in the queue identify the earliest timestep after $\delta_\text{max}$ (in six-hourly outputs) that $R_a(t)>s$ and call this time $t_a^s$. At an \emph{earlier} time $t_a^s-\delta$ (which can be in the initial length-$\delta_\text{max}$ runway), spawn a new ``child'' $\widetilde{X}$ which shares its parent's history up until $t_a^s-\delta$, but then gets perturbed by use of a new seed for random number generation in the stochastic parameterization scheme (or a small random kick if the model is deterministic). Thereafter, the child diverges from its parent for the remainder of the simulation until the time horizon ends at $T$. $\delta$ is the key \emph{advance split time} (AST) parameter, which we vary systematically in this study from 0 to 20 days. Calculate the child's severity $\widetilde{S}$ as the maximum of its intensity $\widetilde{R}(t)$ over $0\leq t\leq T$, the same time horizon as used for the parent, which excludes the initial $\delta_\text{max}$ runway. The next step depends on whether the child's severity exceeds $s$:

    \begin{enumerate}
        \item If the child's severity $\widetilde{S}$ exceeds $s$, we call this ``success'' and officially admit the child into the active population: $X_{a_{A+1}}=\widetilde{X}$, with the same probability weight as its parent, and $S_{a_{A+1}}=\widetilde{S}$. To maintain a constant total probability weight in the active population, adjust all active weights by the same factor: $W_a\leftarrow\frac{A}{A+1}W_a$ for all $a\in\mathcal{A}$. 
        Finally, increment $A$ to $A+1$. 

        \item Otherwise, in case the child's severity fails to exceed $s$ (which might happen, because the split happens before the parent's first threshold crossing; see Fig. 1 in \citeA{Finkel2024bringing}), discard the child completely (formally, set its weight to zero) and move to the next parent in the queue to clone it in the same way.

    \end{enumerate}
    
    Keep cycling through the queue until either the active set is fully replenished to a size $A=N$ (the original population size) with $K$ new successful children, or the total number $M$ of simulations (including ancestors, discarded members, and inactive members) exhausts a pre-determined computational budget, $M=M_{\mathrm{max}}$. For our main experiments with $N=16$, we set $M_{\mathrm{max}}=150$. For $N=32$, we set $M_{\mathrm{max}}=300$. 

    \item\underline{Iteration}: Repeatedly perform step 3 starting with the active population, resulting in a higher level $s$, followed by step 4 on the sub-ensemble exceeding $s$. 

    \item\underline{Termination}: halt the algorithm once the number of severity levels in $\mathcal{S}$ exceeds a pre-set number (in our case, 20), or the total number $M$ of simulations reaches the aforementioned budget $M_{\mathrm{max}}$. 

    \item\underline{Post-analysis}: For any observable of interest expressible as $F(X)$, where $X$ denotes a random variable comprising a whole trajectory $\{X(t):0<t\leq T\}$ with $X(0)$ drawn from $\rho_0$, and $F$ is a generic functional, estimate its expectation as 
    \begin{align}\label{eq:weighted_average}
        \widehat{F}=\frac{\sum_{m=1}^MW_mF(X_m)}{\sum_{m=1}^MW_m}.
    \end{align}
    The denominator is always equal to $N$. In particular, for any given severity $s$, an estimate $\widehat{\mathbb{P}}\{S>s\}$ for its exceedance probability is found by defining $F(X):=\mathbb{I}\{S(X)>s\}$ in the formula above, where $\mathbb{I}$ is the indicator function (one if its argument is true, zero otherwise). The corresponding \emph{return period} $\tau(s)$---the average time between consecutive exceedances, using a Poisson process statistical model---is estimated following \citeA{Lestang2018computing} as 
    \begin{align}\label{eq:return_time_poisson}
        \widehat{\tau}(s)=-\frac{T}{\log[1-\widehat{\mathbb{P}}\{S>s\}]}, 
    \end{align}  
    where $T$ is the time horizon. 
\end{enumerate}
The estimator~\eqref{eq:weighted_average} is unbiased, meaning correct \emph{in expectation} over different independent repetitions (``runs'') of the entire procedure, each of which is itself random. In the results to follow, we have generated 48 independent runs with different ancestors and random seeds to get an accurate sense of variability across runs. The total cost of a single TEAMS run, as reported below in Fig.~\ref{fig:returncurves}, is taken to be $M(\delta+T)$, where $M$ is the total number of members generated in the run ($\leq M_\text{max}$; 150 for $N=16$ and 300 for $N=32$), $\delta$ is the advance split time ($\leq\delta_\text{max}$=25 days) and $T$ is the time horizon ($=35$ days). These costs range from 20-40 years per run in the experiments shown.

This version of TEAMS mostly follows the version in \citeA{Finkel2024bringing}, but differs in two substantial ways. First, in step 4, the previous version of TEAMS would allow parents to stand in for their failed children, and raise the level after $K$ cloning attempts even if they all fail, whereas the new version refuses to raise the level before children alone repopulate the ensemble. Heuristically, the new version is more like mastery-based learning \cite{Winget2022practical}, wherein students only advance after demonstrating mastery even if it takes a longer time with remedial coursework. Even if the levels don't advance as high this way, it ensures that the levels reached are more thoroughly sampled and avoids overextending an ``aging'' ensemble beyond its means. Of course, this risks stagnation at a single level that is impossible to overcome. To cut our losses, we impose a lean budget of $M_\mathrm{max}=150$ (when $N=16$) or $M_\mathrm{max}=300$ (when $N=32$) as the second major difference from \citeA{Finkel2024bringing}, where the budget was 1024 (with $N=128$) and in practice was rarely reached because of a second ``diversity'' criterion that is not used here.   
We have found this version to give more reliable speedup at shorter return periods with reasonable costs, and to reduce  ``apparent bias'': the same phenomenon that causes a coin with true ``tails'' probability of 1/100 to underestimate it at zero on 99\% of flips, even though each flip is unbiased. The previous version of TEAMS had a similar high probability of underestimating return levels in any given run, despite being unbiased, and the algorithmic modification was critical for extending this algorithm from a toy model (Lorenz-96) to a GCM.

We highlight two connections between this new version of TEAMS and other existing algorithms. First, in the sense of repeatedly spawning descendants until success (or computational budget overrun), our new version resembles ``anticipated AMS'' \cite{Rolland2022collapse}. However, in another important sense, anticipated AMS still differs by splitting ancestors when $R_a(t)$ crosses a lower threshold than $s$, rather than at a fixed advance split time. This would not work on precipitation, which rises from zero to peak values more rapidly than ensemble members can diverge; hence, the TEAMS strategy of splitting a fixed time in advance. Second, with particular choices of culling schedule, TEAMS could resemble ensemble boosting with probability estimates as laid out in \citeA{BloinWibe2025estimating} and \citeA{Finkel2025boosting}. Ensemble boosting starts by immediately selecting ancestors exceeding an already-extreme threshold, such as the 0.9 quantile. In TEAMS, one could customize the culling schedule by changing $K$ as the level is raised, for example $K=0.95N$ for the first level and $K=1$ subsequently. One could furthermore halt the level-raising at a single level, and increase the population size to several times the initial $N$ (with proper re-weighting as in step 4a with each new descendant), and this would make TEAMS very much resemble ensemble boosting. We don't make such drastic modifications here, as there is some benefit to raising the level more modestly and giving the moderately extreme ancestors more chances to boost, but it is helpful to view these algorithms as existing on a continuum, related to the tradeoff between exploration and exploitation in optimization methods \cite{Rose2021reinforcement}.

The AST, $\delta$, is a crucial hyperparameter underlying TEAMS which must be chosen in a cheap and reliable way in order to scale TEAMS successfully to realistic GCMs. In section \ref{sec:ensemble_spreading_rate}, we estimate the proposed AST from \citeA{Finkel2024bringing}, namely the time until a perturbed ensemble disperses to a fraction $3/8$ of its saturation dispersion, using a branching procedure. But first, we will present results from TEAMS across a range of ASTs, and at two resolutions, to demonstrate its ability to sample extreme events in the GCM.

\section{Results}\label{sec:results}

\subsection{TEAMS performance}\label{sec:teams_performance}
\begin{figure}
    \centering
    \includegraphics[width=0.99\linewidth,trim={0cm 0cm 17cm 0cm},clip]{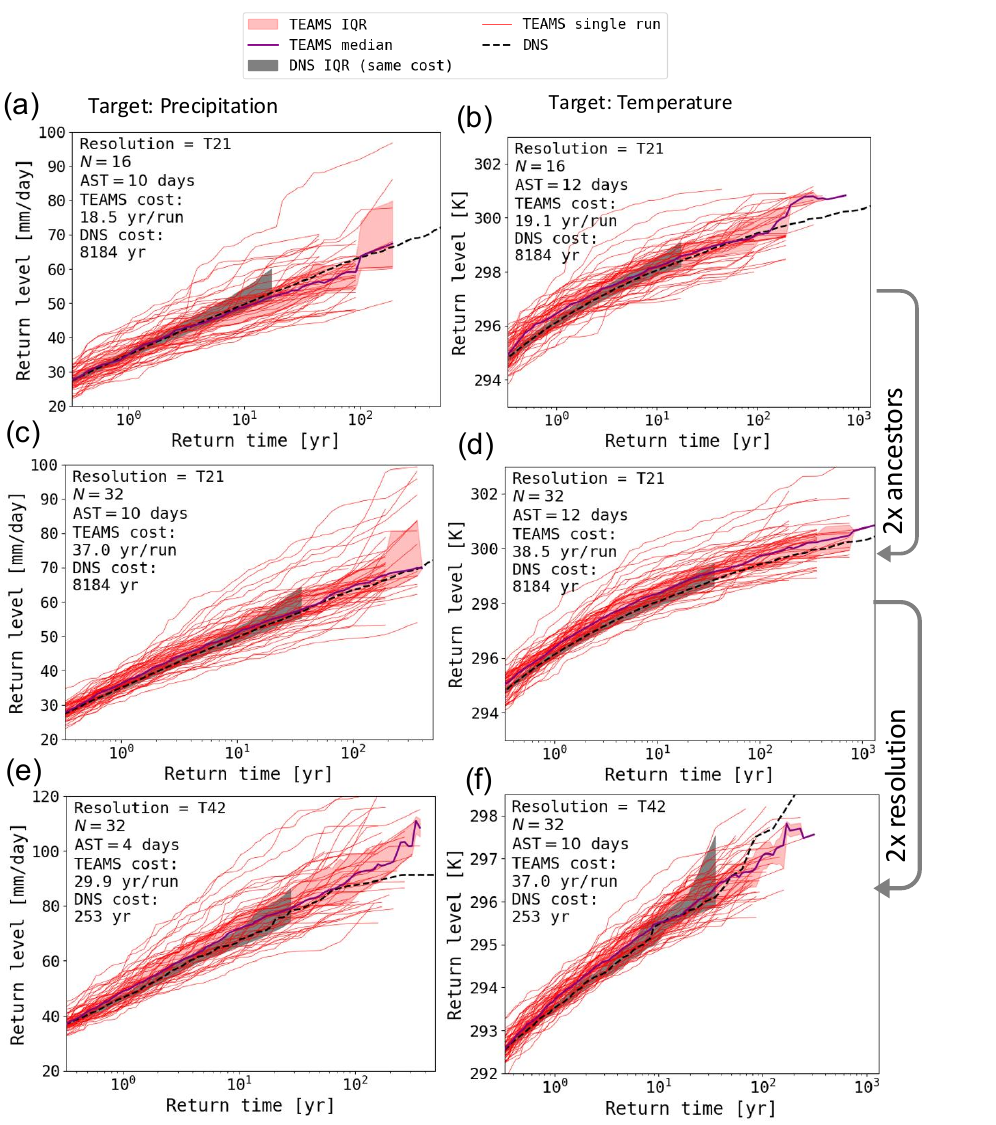}
    \caption{Performance of the rare event algorithm (TEAMS) against the benchmark direct numerical simulation (DNS) in terms of accuracy, uncertainty, and computational cost. Target variables are precipitation (left) and surface temperature (right). Panels (a,b) show the baseline setting: T21 resolution with $N=16$ ancestors. We further perform two ``pivot'' experiments: doubling the ancestors to 32 (c,d) and doubling the resolution to T42 (e,f). Note the difference in vertical axis scale for T42, as resolution can strongly influence the possible ranges. All curves are estimates of return level (event severity) as a function of return period (average inter-event time). Black dashed lines come from a long DNS (``ground truth''). Each thin red line comes from a run of TEAMS with a different random seed (48 in total). Purple lines and red bands indicate medians and inter-quartile ranges (25th-75th percentiles) across the 48 runs, or somewhat fewer in the far tail, restricting to runs with enough splits to estimate the smallest probabilities. For a fair performance comparison, gray error bars show the inter-quartile range of estimates derived from random subsets of the long DNS, equal in cost to a single TEAMS run. Each panel contains a table of corresponding parameters and costs. }
    \label{fig:returncurves}
\end{figure}

In our default configuration of T21 resolution and N=16 ancestors, we ran TEAMS for a range of advance split times $\delta\in\{$0, 4, 6, 8, 10, 12, 14, 16, 20, 24$\}$ days. 
Fig. \ref{fig:returncurves}(a,b) displays the resulting estimates of return level vs. return period for both targets of local precipitation (left), with $\delta=10$ days, and temperature (right), with $\delta=12$ days, which are selected as optimal values based on sensitivity analysis to be presented in Sec. \ref{sec:ast_sensitivity_analysis}. 
In addition, as a test of the algorithm's scaling behavior, we performed two ``pivot'' experiments: doubling the ancestor pool to $N=32$  in Fig. \ref{fig:returncurves}(c,d), and doubling horizontal resolution to T42 in Fig. \ref{fig:returncurves}(e,f). Resolution-doubling tests the algorithm's robustness with more expensive, realistic models, and to what extent lower-resolution versions can inform best practices. To maximize the chances of success at T42 we retained $N=32$ (which proved well worth the extra cost at T21), and re-calibrated the advance split time in light of differing dispersion timescales at higher resolution (see section \ref{sec:ensemble_spreading_rate}).  

Our overall assessment of TEAMS is that it speeds up estimation of extreme events relative to DNS by factors of 5-10. 
Since GCMs are far more expensive than toy models like Lorenz-96, here we focus on the performance of individual runs of TEAMS instead of pooled estimation across many such runs as we did in \citeA{Finkel2024bringing}. 
In Fig. \ref{fig:returncurves}, the median return level across TEAMS runs (purple line) is generally very close to the DNS ground truth (black dashed line), indicating that the overall bias is not large. The red bands in Fig. \ref{fig:returncurves} assess reliability by how close to the ground truth one can expect a single TEAMS run to land with 50\% probability. Clearly, individual runs of TEAMS can deviate substantially from the ground truth---a problem that should abate with larger $N$, but perhaps slowly---making it advisable for practitioners to perform several independent runs to assess uncertainty. However, most runs do cluster near the ground truth, as conveyed by the red error bar.

Comparing red to gray error bars---the latter coming from DNS, computed with a budget equal to a single TEAMS run---we see a tradeoff between accuracy in the bulk and accuracy in the tail of the distribution. 
For the default case of $N=16$ (Fig.\ref{fig:returncurves}a,b), 
one run of TEAMS is equivalent to $\sim19$ years of DNS in computational cost. TEAMS is less certain than DNS in return periods $\lesssim19$ years (the TEAMS computational budget), as indicated by its wider error bars. But TEAMS provides a good estimate for the range $\sim 19-100$ years for precipitation and $\sim 19-150 $ years for temperature, which a $19$-year DNS simply cannot estimate. 
We take the upper range of reliability to be where the error bar starts behaving erratically due to fewer TEAMS runs splitting that many times.
TEAMS performs similarly on precipitation and temperature, even though the tails are shaped quite differently: from extreme value theory, precipitation shape parameters often take both positive and negative signs, indicating unbounded or bounded tails \cite{Ragulina2017generalized}, whereas temperature shape parameters tend to be negative \cite{Krakauer2024it}. 
Extreme value theory could be applied to the DNS to extrapolate return values, but this would not generate dynamical samples of events in the same way that TEAMS does. 
Furthermore, extreme value theory can struggle to capture the shape parameter and hence the far tails correctly, sometimes assigning zero probability to events that are dynamically possible \cite<e.g.,>[]{Zeder2023effect}, which TEAMS can overcome by using the dynamics itself.

Doubling the ancestor pool from $N=16$ to 32 (Fig. \ref{fig:returncurves}c,d) noticeably improves TEAMS' reliability, narrowing the error bars and giving a larger increase in the longest return period. In this case, one TEAMS run is equivalent to just under $40$ years of DNS. We find that one run of TEAMS is less certain than DNS for return periods less than $40$ years, but provides a good estimate for return periods from 40-300 years for precipitation and 40-500 years for temperature, which a $40$ year DNS could not estimate. 

Doubling the resolution  to T42 also leads to good performance and speedup (Fig. \ref{fig:returncurves}e,f) albeit not quite as good as for T21. We chose shorter advance split times for T42 based on ensemble dispersion experiments as discussed in detail in Section \ref{fig:ensemble_dispersion}. The T42 runs are significantly more expensive: besides doubling horizontal resolution, we also increased vertical levels from 6 to 30 and reduced the timestep from 2400 to 600 seconds.
We expect that further experimentation with advance split times and population control parameters (such as $N$, $K$) should make improvements possible at this and much higher resolutions. Although the performance is contingent on hyperparameters, we have demonstrated generalization to higher resolution, which is enough to draw cautious optimism for the algorithm's scalability.

\subsection{Case studies and population dynamics}\label{sec:case_studies}

\begin{figure}
    \centering
    \includegraphics[width=0.98\linewidth,trim={0cm 0cm 0cm 0cm},clip]{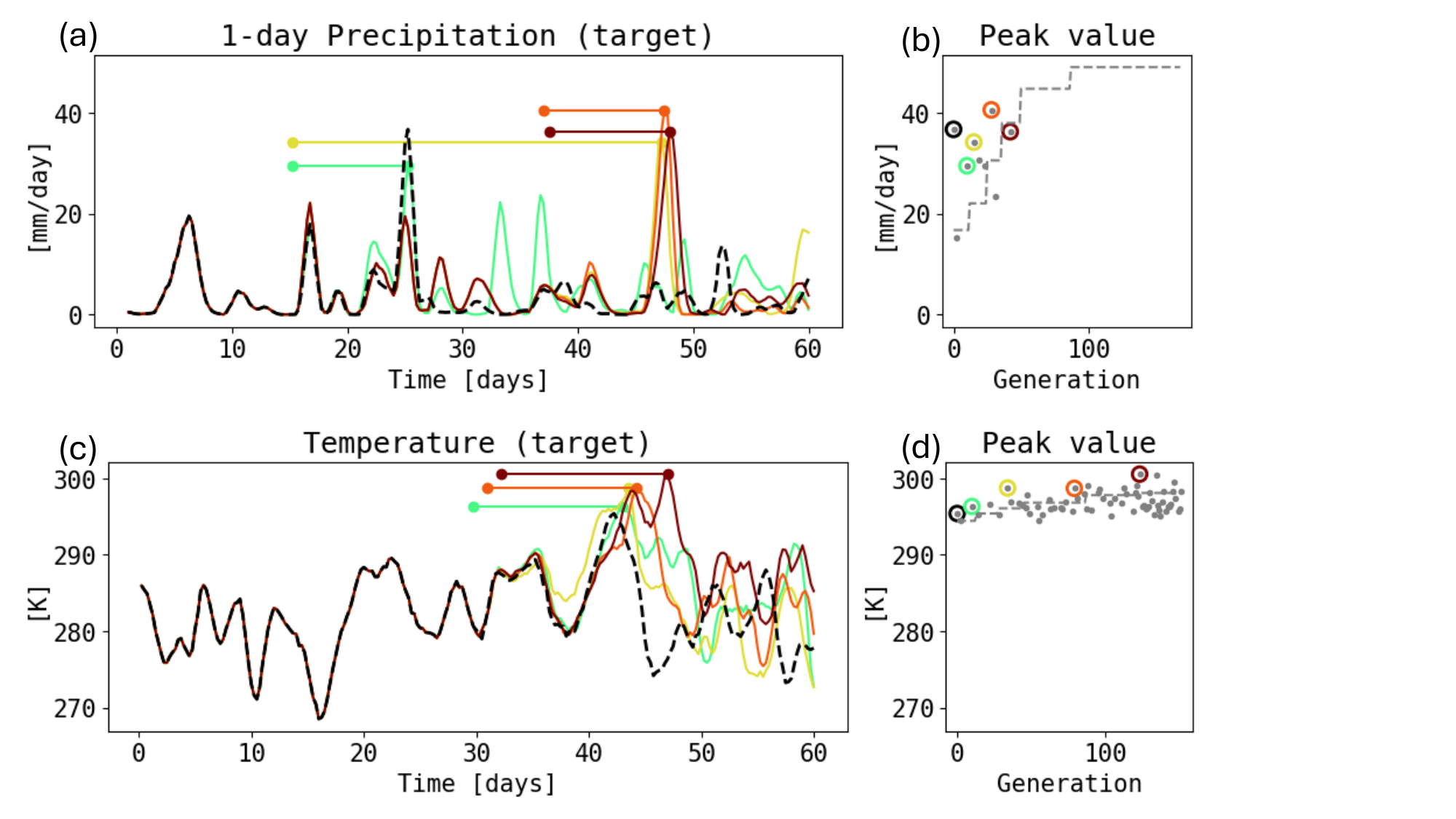}
    \caption{Examples of boosted simulations produced by TEAMS at T21 resolution. Results are shown for (a,b) precipitation with advance split time 10 days, and (c,d) temperature with advance split times 12 days---the values found to be optimal at this resolution. In panels (a,c), black dashed curves are the ancestor and colored curves are descendants (only those in the same lineage as the most-extreme descendant---the ``most-extreme lineage''). Each descendant's split time and peak time are marked by circles connected by a horizontal line (note that orange and yellow lines in 3c overlap). In panels (b,d), the full sequence of descendant severities is shown as gray dots, and those in the most-extreme lineage are also circled in color. Their horizontal position indicates the generation of splitting at which they were spawned, and the dashed gray staircase indicates the algorithm's level $s$ at that same generation. Dots falling below the staircase represent rejections, while those rising above are accepted. There are more gray dots in (d) because the family in (c,d) happened to survive for more rounds of level-raising than the family in (a,b). }
    \label{fig:spaghetti}
\end{figure}

We can better understand the mechanism for TEAMS' success by examining a few case studies, or ``storylines'', of events which are mutated from moderate ancestors into extreme descendants. Fig. \ref{fig:spaghetti} displays one case study for each target variable (precipitation and temperature), with the same advance split times as used at T21 in Fig. \ref{fig:returncurves}(a-d) (10 and 12 days, respectively). Boosting happens either by amplifying an existing spike, or by materializing a new spike where none existed before. In Fig. \ref{fig:spaghetti}a, the first cloning (green) mutated the ancestral spike into a smaller spike, but still cleared the threshold ($\sim20$ mm/day), whereas the second cloning (yellow) first produced an even smaller spike at $t\approx25$ but then discovered a new spike at $t\approx48$. The two subsequent descendants (orange and brown) built further on this second spike, ultimately rising above the ancestor's original score. In Fig. \ref{fig:spaghetti}c, descendants build on the original spike leading to higher and higher severities. Intuitively, this is the more desirable behavior for TEAMS, going by the heuristic guidance that ``the apple shouldn't fall too far from the tree'', or equivalently, subsequent generations should ``stand on the shoulders of their predecessors''. Shortening the time horizon $T$ might help ensure this behavior, but it would limit the discovery of new events that do, in practice, appear to contribute to the best successes of TEAMS so far by allowing later generations to distinguish themselves. We surmise that employing more deterministic optimization strategies, such as Newton's method in the space of perturbations, might help to get the most possible improvement out of existing peaks and obviating the need to get lucky by discovering completely new events. This is a primary direction of our ongoing research. 

\begin{figure}
    \centering
    \includegraphics[width=0.99\linewidth,trim={0cm 0cm 8cm 0cm},clip]{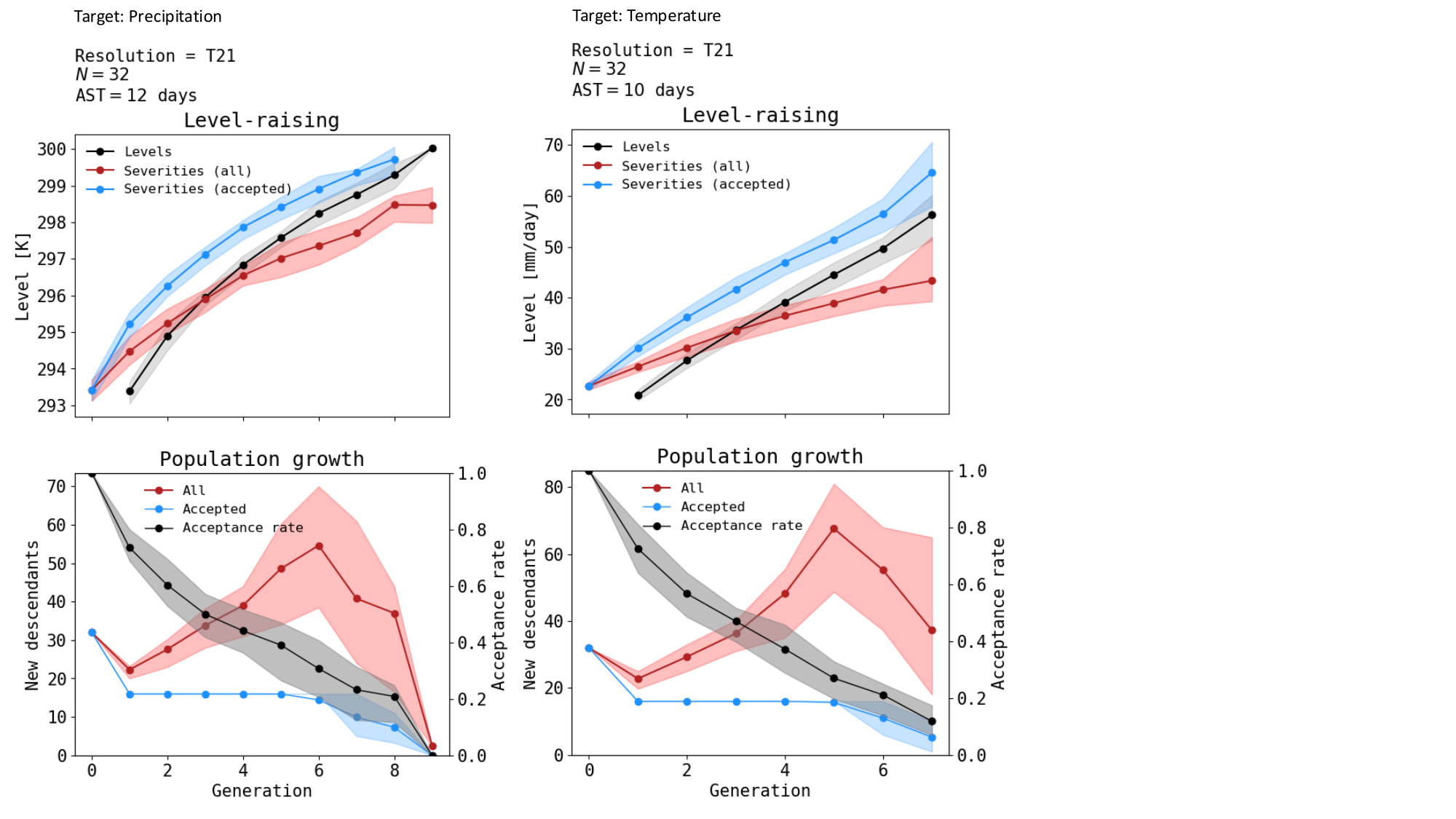}
    \caption{Progression of the population size and severities during TEAMS for both precipitation (a,c) and temperature (b,d) targets at T21 resolution and N=32. (a,b) The levels $s$ (gray), the mean severities across all new descendants born that generation (red), and the mean severities across accepted new descendants only (blue) in a run, as a function of ``generation'' or how many levels have been set so far. Lines and shaded bands show means and interquartile ranges across the 48 runs. (c,d): Population growth per generation (red), accepted population growth (blue), and acceptance rate (black) at each round of level-raising. Population growth accelerates from generations 1-5 and declines thereafter.}
    \label{fig:popstats}
\end{figure}

The ``population dynamics'' of TEAMS offers another window into its behavior and a diagnostic for possible improvements. Fig.~\ref{fig:spaghetti}(b,d) shows aspects of the ensemble members' progress through generations for the same case study.
The level $s$ rises in a stepwise manner while descendant scores rise, on average, only gradually over successive generations, eventually falling systematically below the levels and increasing the rejection rate in later stages of the algorithm. The same story plays out when averaging over runs in Fig.~\ref{fig:popstats}(a,b): severities reached by new ensemble members rise at a slower rate than the levels. The curves cross at generation 3, both for precipitation and temperature despite the differently shaped curves. This coincides with the acceptance rate, shown as the black lines in panels c-d, first dropping below $1/2$.  Total population growth accelerates from generations 1-5 as the algorithm has to try more to produce successful children, and slows down thereafter. We speculate that raising the acceptance rate might improve the overall efficiency and therefore return period.  This might be done by adaptively decreasing the advance split time as the algorithm progresses. More deliberate choices of perturbations might also help to increase acceptance, but with important implications on the assignment of weights. These strategies are beyond our current scope, but we suggest the diagnostics in Fig.~\ref{fig:popstats} as helpful in pursuit of them.

\subsection{Sensitivity analysis of advance split time} \label{sec:ast_sensitivity_analysis}

\begin{figure}
    \centering\includegraphics[width=0.98\linewidth,trim={0cm 3cm 0cm 0cm},clip]{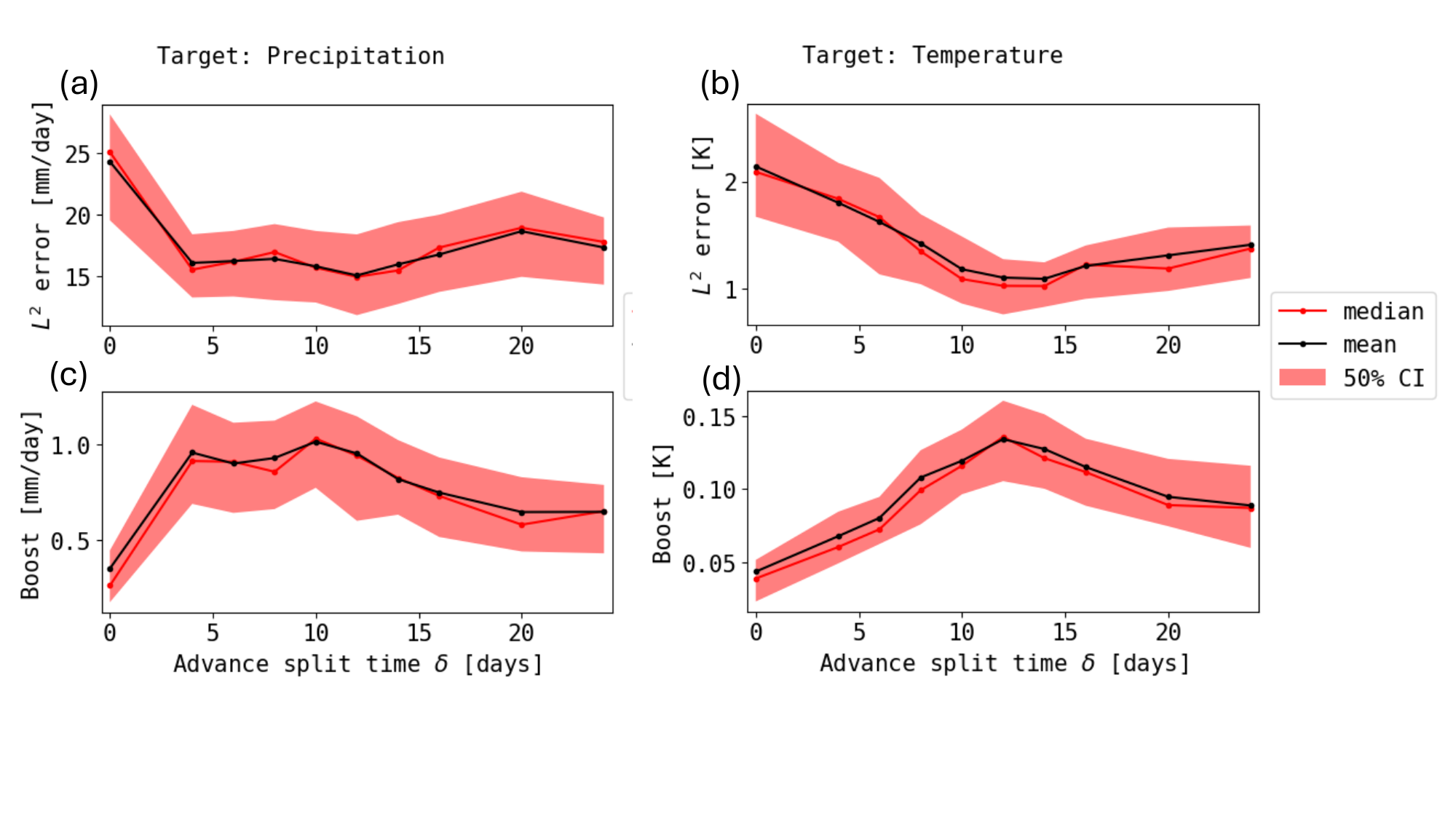}
    \caption{TEAMS performance diagnostics as functions of advance split time for T21 resolution and $N=16$. We deployed TEAMS on two different target variables (left: precipitation and right: temperature) with a sequence of ASTs of 0, 4, 6, 8, 10, 12, 14, 16, 20, and 24. Each case was repeated 48 times with different random seeds. The finer AST spacing of 2 days between 6 and 16 was done after an initial sweep with 4-day spacing to identify a broadly optimal region. Optimality is assessed by the two diagnostics shown: (top) $L^2$ error between TEAMS and DNS return level curves, equivalent to the root-mean-square distance between red and black curves in Fig. \ref{fig:returncurves} (smaller is better); and (bottom) the Boost, defined as the maximum increase in severity between an ensemble member and all of its descendants (or zero if all its descendants are less severe), which is averaged over all members in a TEAMS run. Both $L^2$ and Boost are defined for a single TEAMS run, and there are 48 runs performed at each AST, whose (mean, median, interquartile range) are plotted as (black lines, red lines, and red bands) respectively. }
    \label{fig:ast_dependence}
\end{figure}

Fig. \ref{fig:ast_dependence} quantifies the variation in performance with $\delta$ using two simple performance indicators. The first measures \emph{statistical} accuracy in high return levels:
\begin{align}
    L^2\text{ error}=\bigg(\frac1{\log(\tau_{\text{max}}/\tau_{\text{min}})}\int_{\tau_{\text{min}}}^{\tau_{\text{max}}}\big[\widehat{s}_{\text{DNS}}(\tau)-\widehat{s}_{\text{TEAMS}}(\tau)\big]^2\,d\big[\log\tau\big]\bigg)^{1/2}
\end{align}
where $\tau$ is a return period running from $\tau_{\text{min}}=50$ days to $\tau_{\text{max}}=1.6\times10^4$ years, and $\widehat{s}_{(\text{DNS,TEAMS})}(\tau)$ represents the corresponding severity return level estimated by (DNS, TEAMS) by inverting the estimator $\widehat{\tau}(s)$ in Eq. \eqref{eq:return_time_poisson} with linear (in $\log\tau$ space) interpolation. The integral is approximated by numerical quadrature. Because the DNS is longer than the longest return time estimable by TEAMS (and beyond the range shown in Fig. \ref{fig:returncurves}), we extrapolate $\widehat{s}_{\mathrm{TEAMS}}$ to longer return periods using constant extrapolation, which penalizes runs that get stuck at small boosts and abort at shorter return periods. 
The second indicator measures the efficacy in boosting to larger extremes:
\begin{align}
    \text{Boost}=\frac1M\sum_{m=1}^M\max\{\max(S_\ell-S_m,0):X_\ell\text{ is a descendant of }X_m\}
\end{align}
where $M$ is the total number of ensemble members, including all ancestors and all accepted descendants (but not rejects). 
Fig. \ref{fig:ast_dependence} shows both performance indicators' $\delta$-dependance, and adds to the growing collection of examples \cite{Finkel2024bringing,BloinWibe2025estimating,Finkel2025boosting} demonstrating that \emph{an optimal $\delta$ does exist}, in both senses of minimizing $L^2$ (which has a broad valley) and maximizing Boost (which has a relatively narrow peak). Happily, the same $\delta$ is approximately optimal for both, and $L^2$ is not very sensitive to changes in the value by $\lesssim2$ days. However, the two targets of precipitation and temperature have slightly different optimal $\delta$s of 10 and 12 days respectively, which we will show is consistent with slower ensemble dispersion of temperature in Fig. \ref{fig:ensemble_dispersion}. Thus it appears that the appropriate target time is not universal but rather depends, at least weakly, on the choice of target variable. Heuristically, smoother fields like temperature should take longer to decorrelate, and therefore call for a longer advance split time---at least, when our event of interest is a \emph{single-time maximum} which is the setting where TEAMS is helpful. The next section bears this out quantitatively.

\subsection{Ensemble spreading rate}\label{sec:ensemble_spreading_rate}

\begin{figure}
    \centering
    \includegraphics[width=0.98\linewidth,trim={0cm 0cm 16cm 0cm}, clip]{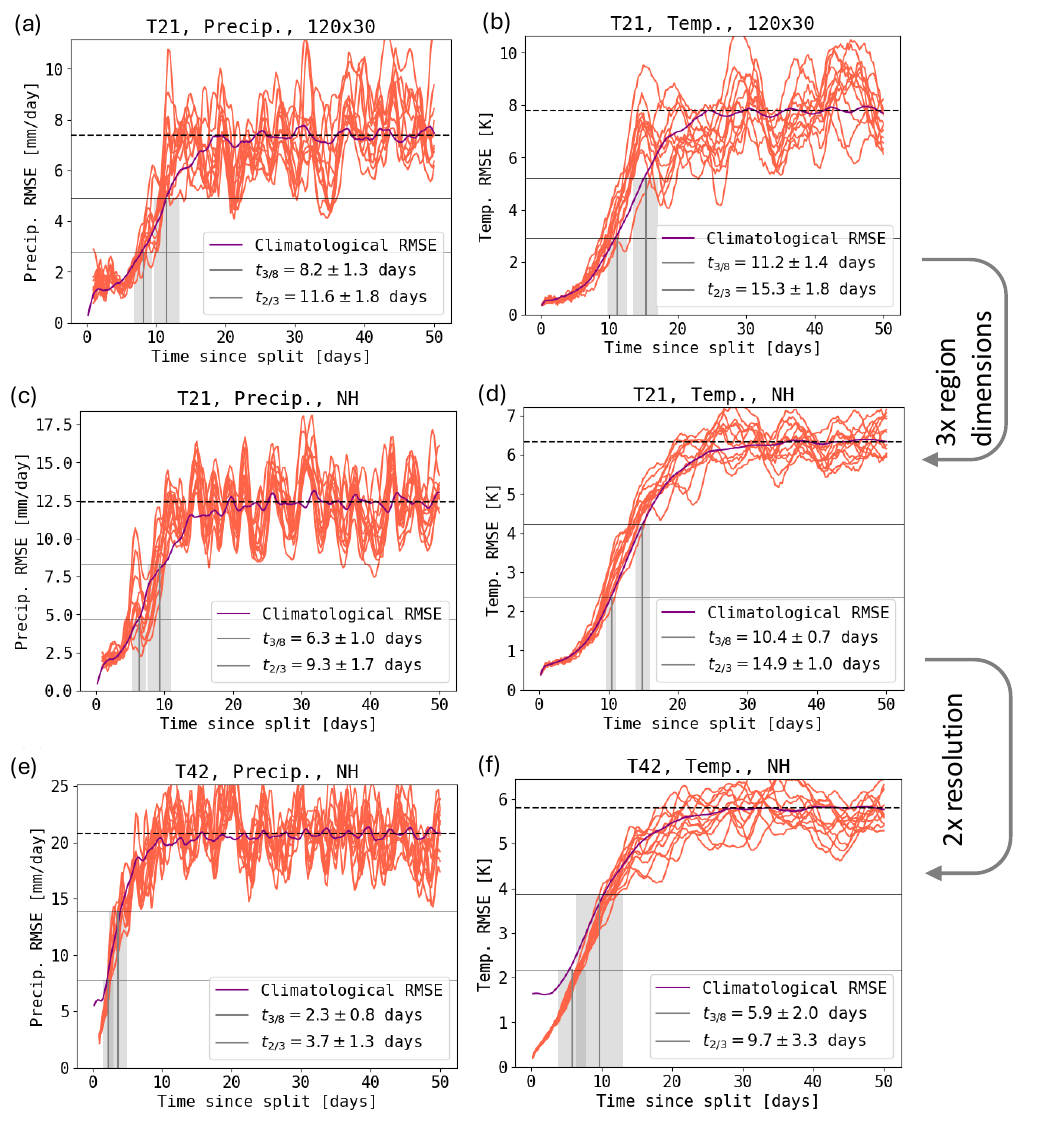}
    \caption{Ensemble dispersion measured using precipitation (a,c,e) and surface temperature (b,d,f) fields.
    Shown in red is the area-weighted Euclidean distance (RMSE) between each realization and the control based on an average over a $120^\circ\times30^\circ$ longitude $\times$ latitude region centered on the target (a,b) or the full northern hemisphere (c-f).
    The RMS of the RMSE over different initial conditions (i.e. different whorls) is shown in purple (denoted RMSE($r$) in the text). The long-term average, or ``saturation RMSE'', is shown as a horizontal black dashed line. The two horizontal gray thresholds mark the fractions $3/8$ and $2/3$ of saturation, and the vertical gray lines with error bars delineate the means and standard deviations of $t_{3/8}$ and $t_{2/3}$, the threshold-crossing time, across whorls as $\overline{t_\epsilon}\pm\mathrm{std}(t_\epsilon)$.  Results are shown for T21 resolution (a-d) and T42 resolution (e,f).}
    \label{fig:ensemble_dispersion}
\end{figure}

\citeA{Finkel2024bringing} found that the optimal $\delta$ was well estimated as the 
time $t_{3/8}$ after which a perturbed ensemble disperses to a fraction $3/8$ of its saturation dispersion. To test the generalization of this rule, we now compare the optimal AST found by grid search in the previous section with $t_{3/8}$ for the GCM, computed by the branching procedure specified below (same as in \citeA{Finkel2024bringing}). The results demonstrate that saturation dispersion remains a useful guide, but the specific value $3/8$ needs quantitative adjustment.  The branching procedure is as follows:

\begin{enumerate}
    \item Draw an initial condition $X(0)\sim\rho_0$, in our case a snapshot from the long DNS run plus some additional spinup of 60 days for good measure.   
    
    \item Split $X(0)$ into $B$ branches (each with its own random seed for SPPT) and let them evolve independently for $T_B$ days. Here we set $B=12$ to balance cost with statistical confidence in estimating root-mean-squared error (RMSE) as defined below. We set $T_B=50$ days which is long enough for the RMSE to saturate. 
    
    \item Continue a simulation from $X(0)$ for an  \emph{equilibration interval} $T_{E}$, and split $X(T_E)$ into $B$ more branches. 
    
    \item Repeat step 3 (but starting from the most recent split time) $W$ times to create $W$ ensembles, resulting in a dataset
    \begin{align}
        \{X_{b,w}(r):1\leq b\leq B, 1\leq w \leq W, 0\leq r\leq T_B\}
    \end{align}
    ($W$ stands for ``whorls'', a botanical term for a point on a stem from which multiple branches emanate). We set $W=20$. $r$ denotes the time since the split, equivalent to $t-(w-1)T_E$ for the $w$th whorl.   

    \item Measure the ensemble dispersion from each whorl $w=1,\hdots,W$ in terms of the RMSE as a function of the elapsed time $r$ since the split:
    \begin{align}
        \mathrm{RMSE}_w(r)=\sqrt{
            \frac1{B}\sum_{b=1}^{B}
                D\Big(X_{w,b}(r), 
                X_{w,0}(r)\Big)^2
        }
    \end{align}
    Here $X_{w,b}$ refers to the $b$th branch from the $w$th whorl, while $b=0$ denotes the ``tree trunk'' which spawns these branches. The distance function $D(X,Y)$ is Euclidean distance in the physical field of interest calculated over a region, chosen here to be either the entire Northern Hemisphere or a smaller region centered on the target location ($120^\circ\times30^\circ$ longitude $\times$ latitude), which might be more relevant to the event of interest but also more noisy. Fig. \ref{fig:ensemble_dispersion} displays the results  in the form of RMSEs, both with the smaller area at T21 (a, b), the full northern hemisphere at T21 (c, d) and the full northern hemisphere at T42 (e, f). Individual branches, plotted in red, show the impact of different stochastic parameterization realizations. 

    \item Because different initial conditions spread at different rates, $\mathrm{RMSE}_w$ might have different shapes for different whorls, but each will eventually saturate to the same asymptotic value. The RMS of $\mathrm{RMSE}_w(r)$ across all $w$s---i.e., $\sqrt{\frac1W\sum_w\mathrm{RMSE}_w^2(r)}$, denoted $\mathrm{RMSE}(r)$---is displayed as purple lines in Fig. \ref{fig:ensemble_dispersion}, and we estimate the asymptotic RMSE by its final 15-day average.
    Define the fractional saturation time $t_{\epsilon,w}$ as the time $r$ at which $\mathrm{RMSE}_w(r)$ reaches a fraction $\epsilon$ of the asymptotic value. Following the prescription from \citeA{Finkel2024bringing},  $\delta$ should be approximated by $\overline{t_{3/8}}:= \frac1W\sum_{w=1}^W{t_{3/8,w}}$. For the GCM, we actually find $\overline{t_{2/3}}$ to be a better guide, and both are marked in Fig.~\ref{fig:ensemble_dispersion}. The order of averaging is important: $\overline{t_\epsilon}$ is not exactly the same as the time that $\mathrm{RMSE}(r)$ crosses $\epsilon\times$(saturation RMSE), but they are practically indistinguishable for the regions considered here. A benefit of averaging times first instead of RMSEs first is that it gives a straightforward estimate of standard deviation of $t_\epsilon$ across $w$s, which is denoted in the legends along with the mean [$t_\epsilon=\overline{t_\epsilon}\pm\mathrm{std}(t_\epsilon)$] for both $\epsilon=3/8$ and $\epsilon=2/3$. 

\end{enumerate}

The first thing to notice is that RMSE saturates more slowly when measured by temperature instead of precipitation, which is consistent with the previous section's result that optimal AST is longer for temperature than for precipitation. This is unsurprising as temperature is generally a smoother field. In addition, the saturation timescales change with the the area chosen for averaging. At T21, going from the full NH to the $120^\circ\times30^\circ$ region, $t_{3/8}$ increases from 6.3 to 8.2 days for precipitation, and from 10.4 to 11.2 days for temperature. The true optimal AST is in fact \emph{longer} than both these values: 10 days for precipitation and 12 days for temperature. We might shrink the averaging region further to make $t_{3/8}$ the right predictor, but this would be \emph{ad hoc} and subject to increasing noise. Instead we adhere to a full-NH notion of distance, which is roughly analogous 
to what we considered in the Lorenz 96 system. This choice gives a new empirical saturation fraction of $\sim2/3$ as a better match for optimal AST. For $t_{2/3}$ we find 9.3 days for precipitation and 14.9 for temperature, which is roughly consistent with the grid search optimal values given the breadth of the valleys in $L^2$ error in Fig. \ref{fig:ast_dependence}. 

It is with this new rule that we deployed TEAMS on the most expensive test case: T42 resolution and $N=32$ ancestors. Fig.~\ref{fig:ensemble_dispersion}(e,f) shows that the dispersion timescale at T42 is well shorter than at T21. We selected $\delta=4$ and 10 days as the approximate $t_{2/3}$ values for precipitation and temperature, respectively, which yielded the results shown in Fig.~\ref{fig:returncurves}(e,f). We take encouragement from the fact that even rough estimates for $\delta$, with rounding, give substantial speedups. 

Clearly, the general problem of optimizing AST is not yet solved. The leading Lyapunov exponent \cite{Cencini2013lyapunov} is a natural first guess for optimal AST, but it only pertains to infinitesimal errors, whereas we aim for finite-amplitude boosts. Furthermore, the Lyapunov exponents' property of being intrinsic to the system actually make them incapable of adjusting to different targets, which we have clearly shown is necessary. 
Note also that ensemble dispersion timescale depends upon the magnitude of stochastic forcing, as shown in \citeA{Finkel2024bringing}, and would also change if we were to use a deterministic model with small one-off kicks instead. Such factors raise doubts on whether Lyapunov exponents are applicable, but ensemble dispersion is at least clearly defined and measurable in all these cases. 
There are theoretical results emerging related to an optimal AST for maximizing chosen combinations of moments of a boosted distribution \cite<appendix B of >[]{BloinWibe2025estimating}, as well as tentative general rules for an optimal AST based on Bayesian optimization \cite{Finkel2025boosting}, but the results here show the need to customize AST for the model, the resolution, and the target, which are important nuances to bear in mind when expanding to other applications, especially those with different spatiotemporal scales such as mesoscale convective systems. Yet our results hint at \emph{scaling relations} between resolution, regions for calculating RMSE, fractional saturation time, and optimal AST. Here we only aim to show that strong speedups are achievable in a GCM, but a more general calibration method is worth pursuing.

\section{Conclusion}\label{sec:conclusion}

Extreme weather events have long been recognized as a major challenge for risk assessment, which motivates the use and development of suitable rare event algorithms: protocols to perturb simulations, over-sample the extremes, and then correct for the statistical bias introduced. The subclass of extremes which are \emph{sudden} and \emph{transient} resist standard rare event algorithms by simply running their course before the perturbations can take effect. We addressed this problem by augmenting a standard algorithm, adaptive multilevel splitting (AMS) with \emph{early perturbations}, resulting in ``trying early AMS'' (TEAMS). After developing the method on the benchmark Lorenz-96 system in \citeA{Finkel2024bringing}, here we have successfully applied the algorithm to a three-dimensional model of the atmosphere's general circulation, extending the estimable range of return periods to $100-150$ years with only $\sim20$ years of simulation and $300-500$ years with only $\sim40$ years of simulation. 

The key hyperparameter of this algorithm is the \emph{advance split time}: how far ahead of time to perturb a simulated extreme event to optimally sample the range of how much more or less severe that event could have been. Exhaustive experiments with Lorenz-96 informed a heuristic rule to set the advance split time based on ensemble dispersion rates \cite{Finkel2024bringing}, and here we found a similar rule gave good performance for this more complex, albeit idealized, atmospheric model. The performance held for two different target variables (heavy precipitation and heat extremes), and two different resolutions (T21 and T42 horizontal triangular spectral truncation). This first evidence of generalizability leads us to conjecture that a similar rule holds in more complex, realistic GCMs. 

There are several wide avenues for advancing this research. An obvious next step is do testing at higher resolution and/or more realistic GCMs or regional climate models. However, algorithmic improvements are still needed for broad application. In particular, we need improved guidance in how to choose the time horizon $T$ and the population control parameters: ancestor pool $N$, killing rate $K$, and computational budget $M_{\text{max}}$.  More interestingly, the appropriate choice of perturbation space is quite open-ended as a general question, especially when stochastic parameterization is not intrinsically a part of the model. Others have conjectured that the perturbation space is inconsequential provided the magnitude is small \cite{Ragone2018computation}, but this remains to be tested, as we are doing in separate ongoing work. Moreover, utilizing \emph{deterministic optimization} to design a more structured sequence of perturbations (in a similar fashion as \citeA{Farazmand2017variational} and \citeA{Sapsis2020output}) may be a route toward more efficient sampling strategies.  

Another immediate goal---beyond our current scope of establishing the TEAMS algorithm, but more and more relevant with more realistic models---is to physically interpret the algorithm's output, which differs from typical datasets in that ensemble members are weighted unequally and grouped into ``families''. Spatial composites of relevant fields, like column water vapor, can be extracted by applying the weighted-average formula \eqref{eq:weighted_average} pointwise to maps, which has been done for seasonal heat extremes in, e.g., \citeA{Ragone2018computation,Ragone2021rare,Miloshevich2024extreme,Lepriol2024using}, and \citeA{Noyelle2025evolution}. In particular, visualizing \emph{differences} between an ancestor and its descendants in this way will reveal mechanisms for physical drivers that strengthen or dampen extremes, and can be compared with traditional perturbations used in numerical weather prediction like Lyapunov, singular, and bred vectors \cite<e.g.,>[]{Norwood2013lyapuNov,Palmer2013singular}. The value added by rare event algorithms is the chance to greatly enhance statistical confidence in composite maps and other diagnostics. For the sake of brevity and to focus on the main point of algorithm development, we leave detailed spatial dynamical analysis to future work.

Overall, we wish to convey simultaneous signals of caution and optimism. ``Extreme weather events'' do not comprise a monolithic category, but are tremendously diverse in spatiotemporal scales, and one rare event algorithm off the shelf cannot be expected to successfully sample all of them. Here we have identified one particular dimension of challenge---relative timescales of ensemble dispersion and the event itself---and successfully remedied it using insight from a simpler model. The specific algorithm, and the general strategy for leveraging a model hierarchy, will help guide the community's continued exploration of extreme events, a growing frontier of climate research.

\section*{Data availability statement}
The code to run the climate model and the rare event algorithm is publicly accessible in two repositories: 
\begin{enumerate}
    \item ``\texttt{jf\_conv\_gray\_smooth}'' \cite<>[available at \url{https://doi.org/10.5281/zenodo.16878347}]{justinfocus122025jfconvgraysmooth} contains the core Fortran model code
    \item ``\texttt{TEAMS}'' \cite<>[available at \url{https://doi.org/10.5281/zenodo.16878339}]{justinfocus122025TEAMS} contains Python code for the rare event algorithm that wraps the Fortran code as well as some other example systems (including Lorenz-96). 
\end{enumerate}   
Interested readers should contact J. F. (\url{ju26596@mit.edu}) for guidance on using and extending the code.

\section*{Conflict of interest}
The authors declare no conflicts of interest relevant to this study.

\acknowledgments
We thank Judith Berner for assistance with implementing the stochastic parameterization. We also extend thanks to three anonymous reviewers for insightful feedback that helped to strengthen the the paper substantially. Computations for this project were performed on the MIT Engaging cluster. This research is part of the MIT Climate Grand Challenge on Weather and Climate Extremes. Support was provided by Schmidt Sciences, LLC.

%
\bibliography{references}

@article{Lestang2018computing,
	doi = {10.1088/1742-5468/aab856},
	url = {https://doi.org/10.1088/1742-5468/aab856},
	year = 2018,
	month = {Apr},
	publisher = {{IOP} Publishing},
	volume = {2018},
	number = {4},
	pages = {043213},
	author = {Thibault Lestang and Francesco Ragone and Charles-Edouard Br{\'{e}}hier and Corentin Herbert and Freddy Bouchet},
	title = {Computing return times or return periods with rare event algorithms},
	journal = {Journal of Statistical Mechanics: Theory and Experiment}
}

@article{Webber2019practical,
author = {Webber,Robert J.  and Plotkin,David A.  and O’Neill,Morgan E  and Abbot,Dorian S.  and Weare,Jonathan },
title = {Practical rare event sampling for extreme mesoscale weather},
journal = {Chaos: An Interdisciplinary Journal of Nonlinear Science},
volume = {29},
number = {5},
pages = {053109},
year = {2019},
doi = {10.1063/1.5081461},

URL = { 
        https://doi.org/10.1063/1.5081461
    
},
eprint = { 
        https://doi.org/10.1063/1.5081461
    
}

}

@article{Abbot2021rare,
doi = {10.3847/1538-4357/ac2fa8},
url = {https://dx.doi.org/10.3847/1538-4357/ac2fa8},
month = {Dec},
year = {2021},
publisher = {The American Astronomical Society},
volume = {923},
number = {2},
pages = {236},
author = {Dorian S. Abbot and Robert J. Webber and Sam Hadden and Darryl Seligman and Jonathan Weare},
title = {Rare Event Sampling Improves Mercury Instability Statistics},
journal = {The Astrophysical Journal}
}

@article {Ragone2018computation,
	author = {Ragone, Francesco and Wouters, Jeroen and Bouchet, Freddy},
	title = {Computation of extreme heat waves in climate models using a large deviation algorithm},
	volume = {115},
	number = {1},
	pages = {24--29},
	year = {2018},
	doi = {10.1073/pnas.1712645115},
	publisher = {National Academy of Sciences},
	issn = {0027-8424},
	URL = {https://www.pnas.org/content/115/1/24},
	eprint = {https://www.pnas.org/content/115/1/24.full.pdf},
	journal = {Proceedings of the National Academy of Sciences}
}

@article {Gessner2021very,
      author = "Claudia Gessner and Erich M. Fischer and Urs Beyerle and Reto Knutti",
      title = "Very Rare Heat Extremes: Quantifying and Understanding Using Ensemble Reinitialization",
      journal = "Journal of Climate",
      year = "2021",
      publisher = "American Meteorological Society",
      address = "Boston MA, USA",
      volume = "34",
      number = "16",
      doi = "10.1175/JCLI-D-20-0916.1",
      pages=      "6619 - 6634",
      url = "https://journals.ametsoc.org/view/journals/clim/34/16/JCLI-D-20-0916.1.xml"
}

@phdthesis{Gessner2022physical,
  title={Physical storylines for very rare climate extremes},
  author={Gessner, Claudia},
  year={2022},
  school={ETH Zurich}
}

@article{Kahn1951estimation,
  title={Estimation of particle transmission by random sampling},
  author={Kahn, Herman and Harris, Theodore E},
  journal={National Bureau of Standards applied mathematics series},
  volume={12},
  pages={27--30},
  year={1951}
}

@article{Zuckerman2017weighted,
author = {Zuckerman, Daniel M. and Chong, Lillian T.},
title = {Weighted Ensemble Simulation: Review of Methodology, Applications, and Software},
journal = {Annual Review of Biophysics},
volume = {46},
number = {1},
pages = {43-57},
year = {2017},
doi = {10.1146/annurev-biophys-070816-033834},
    note ={PMID: 28301772},

URL = { 
        https://doi.org/10.1146/annurev-biophys-070816-033834
    
},
eprint = { 
        https://doi.org/10.1146/annurev-biophys-070816-033834
    
}
}

@article{Ragone2021rare,
author = {Ragone, F. and Bouchet, F.},
title = {Rare Event Algorithm Study of Extreme Warm Summers and Heatwaves Over Europe},
journal = {Geophysical Research Letters},
volume = {48},
number = {12},
pages = {e2020GL091197},
doi = {https://doi.org/10.1029/2020GL091197},
url = {https://agupubs.onlinelibrary.wiley.com/doi/abs/10.1029/2020GL091197},
eprint = {https://agupubs.onlinelibrary.wiley.com/doi/pdf/10.1029/2020GL091197},
note = {e2020GL091197 2020GL091197},
year = {2021}
}

@article{Cerou2007adaptive,
author = { Frédéric   Cérou  and  Arnaud   Guyader },
title = {Adaptive Multilevel Splitting for Rare Event Analysis},
journal = {Stochastic Analysis and Applications},
volume = {25},
number = {2},
pages = {417-443},
year  = {2007},
publisher = {Taylor \& Francis},
doi = {10.1080/07362990601139628},

URL = { 
    
        https://doi.org/10.1080/07362990601139628
    
    

},
eprint = { 
    
        https://doi.org/10.1080/07362990601139628
    
    

}
}

@article{Lestang2020numerical,
title={Numerical study of extreme mechanical force exerted by a turbulent flow on a bluff body by direct and rare-event sampling techniques},
volume={895},
DOI={10.1017/jfm.2020.293},
journal={Journal of Fluid Mechanics},
publisher={Cambridge University Press},
author={Lestang, Thibault and Bouchet, Freddy and Lévêque, Emmanuel},
year={2020},
pages={A19}
}

@article{Au2001estimation,
title = {Estimation of small failure probabilities in high dimensions by subset simulation},
journal = {Probabilistic Engineering Mechanics},
volume = {16},
number = {4},
pages = {263-277},
year = {2001},
issn = {0266-8920},
doi = {https://doi.org/10.1016/S0266-8920(01)00019-4},
url = {https://www.sciencedirect.com/science/article/pii/S0266892001000194},
author = {Siu-Kui Au and James L. Beck}
}

@article {Frierson2006gray,
      author = "Dargan M. W. Frierson and Isaac M. Held and Pablo Zurita-Gotor",
      title = "A Gray-Radiation Aquaplanet Moist GCM. Part I: Static Stability and Eddy Scale",
      journal = "Journal of the Atmospheric Sciences",
      year = "2006",
      publisher = "American Meteorological Society",
      address = "Boston MA, USA",
      volume = "63",
      number = "10",
      doi = "10.1175/JAS3753.1",
      pages=      "2548 - 2566",
      url = "https://journals.ametsoc.org/view/journals/atsc/63/10/jas3753.1.xml"
}

@article {OGorman2008hydrological,
      author = "Paul A. O'Gorman and Tapio Schneider",
      title = "The Hydrological Cycle over a Wide Range of Climates Simulated with an Idealized GCM",
      journal = "Journal of Climate",
      year = "2008",
      publisher = "American Meteorological Society",
      address = "Boston MA, USA",
      volume = "21",
      number = "15",
      doi = "10.1175/2007JCLI2065.1",
      pages=      "3815 - 3832",
      url = "https://journals.ametsoc.org/view/journals/clim/21/15/2007jcli2065.1.xml"
}

@article {OGorman2009scaling,
      author = "Paul A. O’Gorman and Tapio Schneider",
      title = "Scaling of Precipitation Extremes over a Wide Range of Climates Simulated with an Idealized GCM",
      journal = "Journal of Climate",
      year = "2009",
      publisher = "American Meteorological Society",
      address = "Boston MA, USA",
      volume = "22",
      number = "21",
      doi = "10.1175/2009JCLI2701.1",
      pages=      "5676 - 5685",
      url = "https://journals.ametsoc.org/view/journals/clim/22/21/2009jcli2701.1.xml"
}

@article{Wouters2016rare,
doi = {10.1088/1751-8113/49/37/374002},
url = {https://dx.doi.org/10.1088/1751-8113/49/37/374002},
year = {2016},
month = {Aug},
publisher = {IOP Publishing},
volume = {49},
number = {37},
pages = {374002},
author = {J Wouters and F Bouchet},
title = {Rare event computation in deterministic chaotic systems using genealogical particle analysis},
journal = {Journal of Physics A: Mathematical and Theoretical}
}

@article{Huang2016assessing,
title = {Assessing small failure probabilities by AK–SS: An active learning method combining Kriging and Subset Simulation},
journal = {Structural Safety},
volume = {59},
pages = {86-95},
year = {2016},
issn = {0167-4730},
doi = {https://doi.org/10.1016/j.strusafe.2015.12.003},
url = {https://www.sciencedirect.com/science/article/pii/S0167473016000035},
author = {Xiaoxu Huang and Jianqiao Chen and Hongping Zhu}
}

@article{Cencini2013lyapunov,
doi = {10.1088/1751-8113/46/25/250301},
url = {https://dx.doi.org/10.1088/1751-8113/46/25/250301},
year = {2013},
month = {Jun},
publisher = {},
volume = {46},
number = {25},
pages = {250301},
author = {Massimo Cencini and Francesco Ginelli},
title = {Lyapunov analysis: from dynamical systems theory to applications},
journal = {Journal of Physics A: Mathematical and Theoretical},
}

@article{Norwood2013lyapunov,
doi = {10.1088/1751-8113/46/25/254021},
url = {https://dx.doi.org/10.1088/1751-8113/46/25/254021},
year = {2013},
month = {Jun},
publisher = {IOP Publishing},
volume = {46},
number = {25},
pages = {254021},
author = {Adrienne Norwood and Eugenia Kalnay and Kayo Ide and Shu-Chih Yang and Christopher Wolfe},
title = {Lyapunov, singular and bred vectors in a multi-scale system: an empirical exploration of vectors related to instabilities},
journal = {Journal of Physics A: Mathematical and Theoretical}
}

@article{Palmer2013singular,
doi = {10.1088/1751-8113/46/25/254018},
url = {https://dx.doi.org/10.1088/1751-8113/46/25/254018},
year = {2013},
month = {Jun},
publisher = {IOP Publishing},
volume = {46},
number = {25},
pages = {254018},
author = {T N Palmer and Laure Zanna},
title = {Singular vectors, predictability and ensemble forecasting for weather and climate},
journal = {Journal of Physics A: Mathematical and Theoretical}
}

@article{Sapsis2020output,
author = {Sapsis, Themistoklis P. },
title = {Output-weighted optimal sampling for Bayesian regression and rare event statistics using few samples},
journal = {Proceedings of the Royal Society A: Mathematical, Physical and Engineering Sciences},
volume = {476},
number = {2234},
pages = {20190834},
year = {2020},
doi = {10.1098/rspa.2019.0834},

URL = {https://royalsocietypublishing.org/doi/abs/10.1098/rspa.2019.0834},
eprint = {https://royalsocietypublishing.org/doi/pdf/10.1098/rspa.2019.0834}
}

@article{Palmer2009stochastic,
  title={Stochastic parametrization and model uncertainty},
  author={Palmer, Tim N and Buizza, Roberto and Doblas-Reyes, F and Jung, Thomas and Leutbecher, Martin and Shutts, Glenn J and Steinheimer, Martin and Weisheimer, Antje},
  journal={ECMWF Technical Memoranda},
  year={2009},
  publisher={ECMWF Reading, UK}
}

@article{Rose2021reinforcement,
doi = {10.1088/1367-2630/abd7bd},
url = {https://dx.doi.org/10.1088/1367-2630/abd7bd},
year = {2021},
month = {Jan},
publisher = {IOP Publishing},
volume = {23},
number = {1},
pages = {013013},
author = {Dominic C Rose and Jamie F Mair and Juan P Garrahan},
title = {A reinforcement learning approach to rare trajectory sampling},
journal = {New Journal of Physics}
}

@article{Zhang2022koopman,
title = {A Koopman framework for rare event simulation in stochastic differential equations},
journal = {Journal of Computational Physics},
volume = {456},
pages = {111025},
year = {2022},
issn = {0021-9991},
doi = {https://doi.org/10.1016/j.jcp.2022.111025},
url = {https://www.sciencedirect.com/science/article/pii/S0021999122000870},
author = {Benjamin J. Zhang and Tuhin Sahai and Youssef M. Marzouk}
}

@article{Rolland2022collapse,
    title={Collapse of transitional wall turbulence captured using a rare events algorithm},
    volume={931},
    DOI={10.1017/jfm.2021.957},
    journal={Journal of Fluid Mechanics},
    author={Rolland, Joran}, 
    year={2022}, 
    pages={A22}
}

@article {Berner2015increasing,
      author = "J. Berner and K. R. Fossell and S.-Y. Ha and J. P. Hacker and C. Snyder",
      title = "Increasing the Skill of Probabilistic Forecasts: Understanding Performance Improvements from Model-Error Representations",
      journal = "Monthly Weather Review",
      year = "2015",
      publisher = "American Meteorological Society",
      address = "Boston MA, USA",
      volume = "143",
      number = "4",
      doi = "10.1175/MWR-D-14-00091.1",
      pages=      "1295 - 1320",
      url = "https://journals.ametsoc.org/view/journals/mwre/143/4/mwr-d-14-00091.1.xml"
}

@article {Berner2009spectral,
      author = "J. Berner and G. J. Shutts and M. Leutbecher and T. N. Palmer",
      title = "A Spectral Stochastic Kinetic Energy Backscatter Scheme and Its Impact on Flow-Dependent Predictability in the ECMWF Ensemble Prediction System",
      journal = "Journal of the Atmospheric Sciences",
      year = "2009",
      publisher = "American Meteorological Society",
      address = "Boston MA, USA",
      volume = "66",
      number = "3",
      doi = "10.1175/2008JAS2677.1",
      pages=      "603 - 626",
      url = "https://journals.ametsoc.org/view/journals/atsc/66/3/2008jas2677.1.xml"
}

@article{Finkel2024bringing,
author = {Finkel, Justin and O’Gorman, Paul A.},
title = {Bringing Statistics to Storylines: Rare Event Sampling for Sudden, Transient Extreme Events},
journal = {Journal of Advances in Modeling Earth Systems},
volume = {16},
number = {6},
pages = {e2024MS004264},
doi = {https://doi.org/10.1029/2024MS004264},
url = {https://agupubs.onlinelibrary.wiley.com/doi/abs/10.1029/2024MS004264},
eprint = {https://agupubs.onlinelibrary.wiley.com/doi/pdf/10.1029/2024MS004264},
note = {e2024MS004264 2024MS004264},
year = {2024}
}

@article{Winget2022practical,
title = {A Practical Review of Mastery Learning},
journal = {American Journal of Pharmaceutical Education},
volume = {86},
number = {10},
pages = {ajpe8906},
year = {2022},
issn = {0002-9459},
doi = {https://doi.org/10.5688/ajpe8906},
url = {https://www.sciencedirect.com/science/article/pii/S0002945923007386},
author = {Marshall Winget and Adam M. Persky}
}

@Article{Krakauer2024it,
AUTHOR = {Krakauer, Nir Y.},
TITLE = {It Is Normal: The Probability Distribution of Temperature Extremes},
JOURNAL = {Climate},
VOLUME = {12},
YEAR = {2024},
NUMBER = {12},
ARTICLE-NUMBER = {204},
URL = {https://www.mdpi.com/2225-1154/12/12/204},
ISSN = {2225-1154},
DOI = {10.3390/cli12120204}
}

@article{Ragulina2017generalized,
author = {Galina Ragulina and Trond Reitan},
title = {Generalized extreme value shape parameter and its nature for extreme precipitation using long time series and the Bayesian approach},
journal = {Hydrological Sciences Journal},
volume = {62},
number = {6},
pages = {863--879},
year = {2017},
publisher = {Taylor \& Francis},
doi = {10.1080/02626667.2016.1260134},


URL = { 
    
        https://doi.org/10.1080/02626667.2016.1260134
    
    

},
eprint = { 
    
        https://doi.org/10.1080/02626667.2016.1260134
    
    

}

}

@article{Farazmand2017variational,
author = {Mohammad Farazmand  and Themistoklis P. Sapsis },
title = {A variational approach to probing extreme events in turbulent dynamical systems},
journal = {Science Advances},
volume = {3},
number = {9},
pages = {e1701533},
year = {2017},
doi = {10.1126/sciadv.1701533},
URL = {https://www.science.org/doi/abs/10.1126/sciadv.1701533},
eprint = {https://www.science.org/doi/pdf/10.1126/sciadv.1701533}}

@article{Miloshevich2024extreme,
  title={Extreme heat wave sampling and prediction with analog Markov chain and comparisons with deep learning},
  author={Miloshevich, George and Lucente, Dario and Yiou, Pascal and Bouchet, Freddy},
  journal={Environmental Data Science},
  volume={3},
  pages={e9},
  year={2024},
  publisher={Cambridge University Press}
}

@article{Sillmann2017understanding,
title = {Understanding, modeling and predicting weather and climate extremes: Challenges and opportunities},
journal = {Weather and Climate Extremes},
volume = {18},
pages = {65-74},
year = {2017},
issn = {2212-0947},
doi = {https://doi.org/10.1016/j.wace.2017.10.003},
url = {https://www.sciencedirect.com/science/article/pii/S2212094717300440},
author = {Jana Sillmann and Thordis Thorarinsdottir and Noel Keenlyside and Nathalie Schaller and Lisa V. Alexander and Gabriele Hegerl and Sonia I. Seneviratne and Robert Vautard and Xuebin Zhang and Francis W. Zwiers}
}

@article{Lepriol2024using,
doi = {10.1088/2752-5295/ad8027},
url = {https://dx.doi.org/10.1088/2752-5295/ad8027},
year = {2024},
month = {oct},
publisher = {IOP Publishing},
volume = {3},
number = {4},
pages = {045016},
author = {Le Priol, Clément and Monteiro, Joy Merwin and Bouchet, Freddy},
title = {Using rare event algorithms to understand the statistics and dynamics of extreme heatwave seasons in South Asia},
journal = {Environmental Research: Climate}
}

@article{Uribe2021cross,
author = {Uribe, Felipe and Papaioannou, Iason and Marzouk, Youssef M. and Straub, Daniel},
title = {Cross-Entropy-Based Importance Sampling with Failure-Informed Dimension Reduction for Rare Event Simulation},
journal = {SIAM/ASA Journal on Uncertainty Quantification},
volume = {9},
number = {2},
pages = {818-847},
year = {2021},
doi = {10.1137/20M1344585},

URL = { 
    
        https://doi.org/10.1137/20M1344585
    
    

},
eprint = { 
    
        https://doi.org/10.1137/20M1344585
    
    

}
}

@misc{Mahesh2024hensone,
      title={Huge Ensembles Part I: Design of Ensemble Weather Forecasts using Spherical Fourier Neural Operators}, 
      author={Ankur Mahesh and William Collins and Boris Bonev and Noah Brenowitz and Yair Cohen and Joshua Elms and Peter Harrington and Karthik Kashinath and Thorsten Kurth and Joshua North and Travis OBrien and Michael Pritchard and David Pruitt and Mark Risser and Shashank Subramanian and Jared Willard},
      year={2024},
      eprint={2408.03100},
      archivePrefix={arXiv},
      primaryClass={physics.ao-ph},
      url={https://arxiv.org/abs/2408.03100}, 
}

@misc{Mahesh2024henstwo,
      title={Huge Ensembles Part II: Properties of a Huge Ensemble of Hindcasts Generated with Spherical Fourier Neural Operators}, 
      author={Ankur Mahesh and William Collins and Boris Bonev and Noah Brenowitz and Yair Cohen and Peter Harrington and Karthik Kashinath and Thorsten Kurth and Joshua North and Travis OBrien and Michael Pritchard and David Pruitt and Mark Risser and Shashank Subramanian and Jared Willard},
      year={2024},
      eprint={2408.01581},
      archivePrefix={arXiv},
      primaryClass={cs.LG},
      url={https://arxiv.org/abs/2408.01581}, 
}

@Article{BloinWibe2025estimating,
AUTHOR = {Bloin-Wibe, L. and Noyelle, R. and Humphrey, V. and Beyerle, U. and Knutti, R. and Fischer, E.},
TITLE = {Estimating return periods for extreme events in climate models through Ensemble Boosting},
JOURNAL = {EGUsphere},
VOLUME = {2025},
YEAR = {2025},
PAGES = {1--40},
URL = {https://egusphere.copernicus.org/preprints/2025/egusphere-2025-525/},
DOI = {10.5194/egusphere-2025-525}
}

@misc{Finkel2025boosting,
      title={Boosting Ensembles for Statistics of Tails at Conditionally Optimal Advance Split Times}, 
      author={Justin Finkel and Paul A. O'Gorman},
      year={2025},
      eprint={2507.22310},
      archivePrefix={arXiv},
      primaryClass={physics.ao-ph},
      url={https://arxiv.org/abs/2507.22310}, 
}

@article {Held2005gap,
      author = "Isaac M. Held",
      title = "The Gap between Simulation and Understanding in Climate Modeling",
      journal = "Bulletin of the American Meteorological Society",
      year = "2005",
      publisher = "American Meteorological Society",
      address = "Boston MA, USA",
      volume = "86",
      number = "11",
      doi = "10.1175/BAMS-86-11-1609",
      pages=      "1609 - 1614",
      url = "https://journals.ametsoc.org/view/journals/bams/86/11/bams-86-11-1609.xml"
}

@article{Anderson2004new,
    author = "Jeffrey L. Anderson and others",
    title = "The New GFDL Global Atmosphere and Land Model AM2–LM2: Evaluation with Prescribed SST Simulations",
    journal = "Journal of Climate",
    year = "2004",
    publisher = "American Meteorological Society",
    address = "Boston MA, USA",
    volume = "17",
    number = "24",
    doi = "10.1175/JCLI-3223.1",
    pages=      "4641 - 4673",
    url = "https://journals.ametsoc.org/view/journals/clim/17/24/jcli-3223.1.xml"
}

@software{justinfocus122025TEAMS,
  author       = {justinfocus12},
  title        = {justinfocus12/TEAMS: Initial release for submission},
  month        = {aug},
  year         = {2025},
  publisher    = {Zenodo},
  version      = {v0.1},
  doi          = {10.5281/zenodo.16878339},
  url          = {https://doi.org/10.5281/zenodo.16878339},
}

@software{justinfocus122025jfconvgraysmooth,
  author       = {justinfocus12},
  title        = {justinfocus12/jf\_conv\_gray\_smooth: Initial releasefor submission},
  month        = {aug},
  year         = {2025},
  publisher    = {Zenodo},
  version      = {v0.1},
  doi          = {10.5281/zenodo.16878347},
  url          = {https://doi.org/10.5281/zenodo.16878347},
}

@article {Tagle2016temperature,
      author = "Felipe Tagle and Judith Berner and Mircea D. Grigoriu and Natalie M. Mahowald and Gennady Samorodnitsky",
      title = "Temperature Extremes in the Community Atmosphere Model with Stochastic Parameterizations",
      journal = "Journal of Climate",
      year = "2016",
      publisher = "American Meteorological Society",
      address = "Boston MA, USA",
      volume = "29",
      number = "1",
      doi = "10.1175/JCLI-D-15-0314.1",
      pages=      "241 - 258",
      url = "https://journals.ametsoc.org/view/journals/clim/29/1/jcli-d-15-0314.1.xml"
}

@book{Krishnamurti2006introduction,
  title={An introduction to global spectral modeling},
  author={Krishnamurti, Tiruvalam Natarajan and Hardiker, VM and Bedi, HS and Ramaswamy, L},
  year={2006},
  publisher={Springer}
}

@article{Zeder2023effect,
author = {Zeder, Joel and Sippel, Sebastian and Pasche, Olivier C. and Engelke, Sebastian and Fischer, Erich M.},
title = {The Effect of a Short Observational Record on the Statistics of Temperature Extremes},
journal = {Geophysical Research Letters},
volume = {50},
number = {16},
pages = {e2023GL104090},
doi = {https://doi.org/10.1029/2023GL104090},
url = {https://agupubs.onlinelibrary.wiley.com/doi/abs/10.1029/2023GL104090},
eprint = {https://agupubs.onlinelibrary.wiley.com/doi/pdf/10.1029/2023GL104090},
note = {e2023GL104090 2023GL104090},
year = {2023}
}

@article{Noyelle2025evolution,
author = {Noyelle, Robin and Caubel, Arnaud and Meurdesoif, Yann and Faranda, Davide and Yiou, Pascal},
title = {Evolution of the Dynamics of Centennial Hot Summers in Western Europe With Climate Change},
journal = {Geophysical Research Letters},
volume = {52},
number = {14},
pages = {e2025GL115552},
doi = {https://doi.org/10.1029/2025GL115552},
url = {https://agupubs.onlinelibrary.wiley.com/doi/abs/10.1029/2025GL115552},
eprint = {https://agupubs.onlinelibrary.wiley.com/doi/pdf/10.1029/2025GL115552},
note = {e2025GL115552 2025GL115552},
year = {2025}
}
%


%
%
%
%
%

\end{document}